\documentclass[fleqn,usenatbib]{mnras}

\usepackage{newtxtext,newtxmath}

\usepackage[T1]{fontenc}

\DeclareRobustCommand{\VAN}[3]{#2}
\let\VANthebibliography\thebibliography
\def\thebibliography{\DeclareRobustCommand{\VAN}[3]{##3}\VANthebibliography}

\usepackage{CJK}

\usepackage{booktabs}
\usepackage{pifont}
\usepackage{xcolor}
\usepackage{bm}
\usepackage{graphicx}
\usepackage[normalem]{ulem}
\usepackage{float}
\usepackage{algorithm}
\usepackage{algorithmic}
\usepackage{listings}
\usepackage{hyperref}
\usepackage{amsmath}	

\newcommand{\teff}{T_{\rm eff}}
\newcommand{\logg}{\log g}
\newcommand{\cmark}{\ding{51}}%
\newcommand{\xmark}{\ding{55}}%

\title[Astroconformer]{Astroconformer: The Prospects of Analyzing Stellar Light Curves \\ with Transformer-Based Deep Learning Models}

\author[Pan et al.]{
Jia-Shu Pan (潘嘉书),$^{1,2}$\thanks{E-mail: jspan@smail.nju.edu.cn}
Yuan-Sen Ting (丁源森),$^{1,3,4}$
Jie Yu (余杰)$^{5,6,3}$\thanks{E-mail: Jie.yu@anu.edu.au}
\\
$^{1}$Research School of Astronomy \& Astrophysics, Australian National University, Cotter Rd., Weston, ACT 2611, Australia∗\\
$^{2}$School of Astronomy and Space Science, Nanjing University, Nanjing 210093, China\\
$^{3}$School of Computing, Australian National University, Acton, ACT 2601, Australia\\
$^{4}$Department of Astronomy, The Ohio State University, Columbus, USA\\
$^{5}$Max Planck Institute for Solar System Research, Justus-von-Liebig-Weg 3, 37077 Göttingen, Germany\\
$^{6}$Heidelberg Institute for Theoretical Studies (HITS) gGmbH, Schloss-Wolfsbrunnenweg 35, D-69118 Heidelberg, Germany
}

\date{Accepted XXX. Received YYY; in original form ZZZ}

\pubyear{2023}

\begin{document}
\begin{CJK*}{UTF8}{gbsn}
\label{firstpage}
\pagerange{\pageref{firstpage}--\pageref{lastpage}}
\maketitle

\begin{abstract}
Stellar light curves contain valuable information about oscillations and granulation, offering insights into stars' internal structures and evolutionary states. Traditional asteroseismic techniques, primarily focused on power spectral analysis, often overlook the crucial phase information in these light curves. Addressing this gap, recent machine learning applications, particularly those using Convolutional Neural Networks (CNNs), have made strides in inferring stellar properties from light curves. However, CNNs are limited by their localized feature extraction capabilities. In response, we introduce \textit{Astroconformer}, a Transformer-based deep learning framework, specifically designed to capture long-range dependencies in stellar light curves. Our empirical analysis centers on estimating surface gravity ($\logg$), using a dataset derived from single-quarter Kepler light curves with $\logg$ values ranging from 0.2 to 4.4. Astroconformer demonstrates superior performance, achieving a root-mean-square-error (RMSE) of 0.017 dex at $\logg\approx3$ in data-rich regimes and up to 0.1 dex in sparser areas. This performance surpasses both K-nearest neighbor models and advanced CNNs. Ablation studies highlight the influence of receptive field size on model effectiveness, with larger fields correlating to improved results. \textit{Astroconformer} also excels in extracting $\nu_{\max}$ with high precision. It achieves less than 2\% relative median absolute error for 90-day red giant light curves. Notably, the error remains under 3\% for 30-day light curves, whose oscillations are undetectable by a conventional pipeline in 30\% cases. Furthermore, the attention mechanisms in \textit{Astroconformer} align closely with the characteristics of stellar oscillations and granulation observed in light curves.
\end{abstract}




\section{Introduction}

\label{sec:intro}

Stellar light curves serve as a comprehensive source for deciphering the diverse physical properties of stars, being shaped by multiple factors such as rotation, spot modulation, oscillations, and granulation \citep{astero1}. Asteroseismology, the study of stellar oscillations, remains a benchmark approach for understanding essential stellar properties like mass, radius, and luminosity \citep{astero1}. Particularly for evolved stars with solar-like oscillations, asteroseismology offers crucial insights into their evolutionary states \citep{astero_evo}, internal rotation \citep{astero_rotation1, astero_rotation2,astero_rotation3}, and magnetic fields \citep{astero_magnetic1, astero_magnetic2, astero_magnetic3}. 

The field of asteroseismology has experienced a significant boost over the past two decades, largely owing to the successful CoRoT \citep{corot} and \textit{Kepler} \citep{Kepler} space missions. The original \textit{Kepler} mission provided high-quality light curves for nearly 200,000 low- to intermediate-mass stars with visual magnitudes below 16, accumulating data over four years in a fixed northern sky field. Subsequent to \textit{Kepler}, the K2 mission \citep{K2} provided an extension, focusing on various ecliptic fields and capturing valuable data for a broad spectrum of stars, albeit within a reduced observational window of approximately 90 days per field. The Transiting Exoplanet Survey Satellite (TESS) \citep{tess}, as \textit{Kepler}'s successor, aims to monitor bright stars throughout the sky. While TESS data holds its own promise for advancing asteroseismology, it poses new challenges due to more limited continuous observation periods, ranging from 27 days to about a year. 

Traditionally, asteroseismic analyses transform light curves into power spectra and analyze summary statistics within this framework \citep{astero1, astero_giant, astero_solar, astero_interior}. Two global seismic parameters, \( \nu_{\max} \) and \( \Delta \nu \), are frequently employed to estimate the parameters of solar-like oscillators. The frequency of maximal oscillation power, \( \nu_{\max} \), is proportional to the acoustic cutoff frequency and is therefore linked to both surface gravity (\( \logg \)) and effective temperature (\(\teff\))  \citep{astero_pmode, astero_numax}. The average frequency separation between consecutive radial order modes at a specific spherical degree, \( \Delta \nu \), scales with mean stellar density as it correlates with the sound travel time through the star \citep{astero_dn}.

However, focusing solely on these oscillation signals within the frequency domain has its drawbacks. For instance, oscillation amplitude is positively correlated with stellar luminosity, making the detection of oscillations in dwarf stars particularly challenging. Moreover, the short-period oscillations observed in main-sequence and subgiant stars necessitate high-cadence photometric observations, which is not widely accessible even for advanced missions like \textit{Kepler} and TESS. Furthermore, not all giants display solar-like oscillations, leaving their correlational relationships with stellar properties yet to be fully understood \citep{non-solar}. Even when applied to giants exhibiting oscillations, conventional pipelines require long observation duration to detect robust oscillations and ascertain global parameters precisely \citep{Hekker_2012}. Furthermore, existing automated pipelines frequently require manual visual inspection—a task that becomes impractical considering the vast volume of data generated by missions like TESS.

The limitations become more pronounced when considering light curves from ground-based observatories. The upcoming US flagship Rubin Observatory \citep{lsst} is slated to commence its sky survey in the next year, adopting a cadence of several days \citep{lsst_strategy}. Similarly, the Zwicky Transient Facility follows a comparable 3-day cadence \citep{ztf}. Such long-cadence observational strategies often exceed the typical timescales of stellar oscillations,  except for long period variables \citep{Yu20}, thereby rendering the detection of oscillations less viable. These considerations, along with the challenges posed by high-noise observations like those encountered in TESS data and varying observational cadences, underline the need for innovative methodologies for robust asteroseismic analysis.

Stellar light curves offer information that extends beyond oscillations and can still be harnessed within the Nyquist limit, even in long-cadence surveys like those of the Rubin Observatory. For instance, stellar granulation offers an alternative avenue for characterizing stars using only long-cadence data \citep{granulation_acoustic,logg_timescale}. Emerging from the convection layer, granulation power correlates with stellar mass and provides insights into a star's evolutionary stage \citep{granulation_acoustic}. Previous studies have quantified the scaling relationship between granulation and global oscillation parameters \citep{granulation_oscillation}. Notably, granulation consists signal on a broader range of timescales compared to oscillations. Consequently, it has been suggested as a more effective means of studying the properties of subgiants and dwarfs \citep{fliper}. In other words, light curves contain subtle features that, if fully exploited, can reveal additional aspects of stellar characteristics.

In light of this, in this study, we introduce a Transformer-based deep learning model, named \textit{Astroconformer}, to analyze stellar light curves. Originating in natural language processing (NLP), the Transformer model utilizes a self-attention mechanism to calculate the correlations—or `attention'—across all inputs in a sequence \citep{attentionisallyouneed}. 
We will demonstrate that Transformer-based methods can more effectively extract essential information from stellar light curves compared to other machine learning approaches. As a proof of concept, we will focus on inferring \( \logg \) and \( \nu_{\max} \) of stars from their light curves.

\section{Relevant Studies and Motivation}
\label{sec:motivation}

In recent years, the field of asteroseismology has increasingly turned to machine learning techniques to glean additional physical insights from stellar light curves, which contain more than just information from stellar oscillations. For instance, \citet{ness2018} used polynomial ridge regression on the autocorrelation function of light curves to attain a \(\logg\) \(<0.1\) dex precision for red giants with \( \logg \) values between \(2.0\) and \(3.5\) dex. Building upon this work, {\sc the swan} \citep{swan} employed a \(k\)-Nearest Neighbors (\(k\)-NN) approach on the power spectra of long-cadence \textit{Kepler} light curves.

Convolutional Neural Networks (CNNs) have been provedalso been utilized in asteroseismology, particularly for their proficiency in tasks such as image recognition \citep{googlenet}. For example, \citet{Hon2018} employed CNNs to scrutinize images of folded power spectra, and successfully classified the evolutionary stages of stars with high accuracy. Moreover, \citet{cnn_lc} employed a one-dimensional CNN to single-quarter {\it Kepler} light curves and recovered $\logg$ of red giants to 0.06 dex, using asteroseismic labels from \citet{yu18}. Recurrent Neural Networks (RNNs), known for their sequence modelling ability, have also been implemented for {\it Kepler} light curves \citep{rnn_lc}.

\begin{figure}
	\includegraphics[width=\columnwidth]{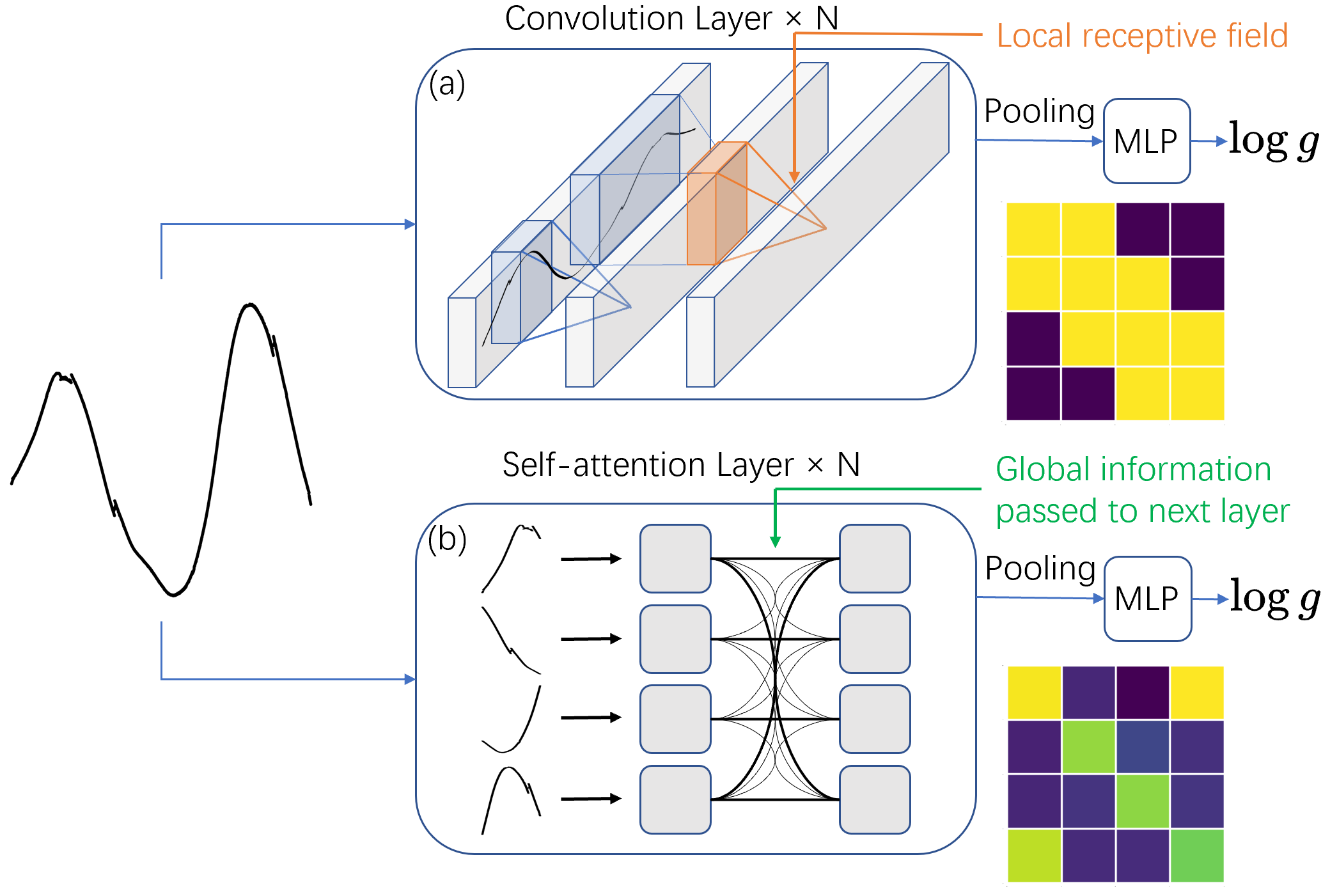}
    \caption{The figure contrasts the limited receptive field of CNNs with the global receptive field afforded by self-attention mechanisms, especially in the context of long sequences like 4-year \textit{Kepler} light curves. (a) In CNNs, multiple convolutional layers with small receptive fields are stacked to achieve a broader receptive field in deeper layers. The heatmap illustrates the pairwise correlations between segments, showing that each segment primarily incorporates information only from its immediate neighbors. (b) Self-attention, on the other hand, calculates the correlations between all segments of the light curve and incorporates this global information into its output.}
    \label{fig:intro_cnn}
\end{figure}

Nevertheless, these data-driven methodologies come with their own sets of limitations in both data handling and modeling. For example, the \(k\)-NN technique used by {\sc the swan} relies on $k$ nearest neighbors for inference. This approach may face challenges in terms of generalization when dealing with sparser regions of the sample space. While deep learning methods typically exhibit better generalization capabilities by adaptively learning features specific to the task at hand \citep{generalization}, the deep learning architectures so far adopted in asteroseismology, such as CNNs and RNNs, also have inherent limitations. Specifically, CNNs are limited by their local receptive fields, capturing only localized spatial information. RNNs, on the other hand, find it challenging to maintain information over long sequences. To mitigate the issue in CNNs, multiple convolutional layers are required to capture interactions between long-range information. This limitation of CNNs is illustrated in Fig.~\ref{fig:intro_cnn}.

Much of the existing analysis has focused on the power spectra of light curves. This focus partially arises because CNNs excel at extracting local information, making them more suited for operating in the frequency domain rather than capturing long-range correlations in time series data. However, relying on power spectra can lead to information loss, particularly when the underlying process is not Gaussian. Power spectra discard the complex phases of the data, which contain important frequency correlations \citep{gaussian1, gaussian2, gaussian3}. Since solar-like oscillations are stochastically excited and intrinsically damped by surface turbulent convection \citep{astero_damp}, they may include subtle non-Gaussian signals.

These limitations—namely, the challenge of directly handling time series data in asteroseismology and the absence of appropriate inductive biases in CNNs for capturing long-range correlations—have led us to investigate Transformer-based models. As we will discuss in subsequent sections, Transformers are inherently designed to capture long-range correlations, making them a more suitable choice for analyzing time series data. Utilizing Transformers allows us to work directly with observed light curves in the time domain, minimizing information loss and obviating the need for additional post-processing steps.

%
%
%
%
%
%

\section{Astroconformer: A Transformer-Based Method to Analyze Stellar Light Curves}
\label{sec:method}

In this section, we introduce \textit{Astroconformer}, a model inspired by the Conformer architecture as presented in \citet{conformer}. Given that the timescale of stellar oscillations and granulation can vary from minutes for main-sequence stars to hundreds of days for most luminous red giants, depending on surface gravity, an effective inference model must capture both local and global features for accurate $\logg$ prediction from light curves. The Conformer model integrates a variety of deep learning modules, combining self-attention layers for long-range information capture with convolutional layers for local feature extraction. Our PyTorch implementation of \textit{Astroconformer} is publicly available\footnote{https://github.com/panjiashu/Astroconformer}. 

\subsection{Self-attention mechanism}
\label{sec:selfattn}

Sequence representation continues to be a vibrant area of research in deep learning, with applications spanning natural language processing (NLP), computer vision (CV) \citep{vit}, and time-series forecasting \citep{patchtst}. Unlike images, where key information is often local, sequences frequently require the extraction of long-range correlations for a comprehensive interpretation.

Traditional CNNs are notably limited in capturing these long-range dependencies effectively, as shown in Fig.~\ref{fig:intro_cnn}. The self-attention mechanism, first introduced by \citet{attentionisallyouneed}, addresses this limitation. Thanks to its remarkable ability to model long-range correlations, self-attention has become a cornerstone in state-of-the-art NLP models like BERT \citep{bert} and GPT-3 \citep{gpt3}.

The crux of self-attention lies in first analyzing the correlation between timestamps (up to some linear combination) and then using this correlation to either amplify or diminish the contributions from individual timestamps. Given that the correlation matrix is calculated across all timestamps, the model inherently incorporates a strong inductive bias towards considering these correlations.

\begin{figure*}
    \centering
    \includegraphics[width=1.0\textwidth]{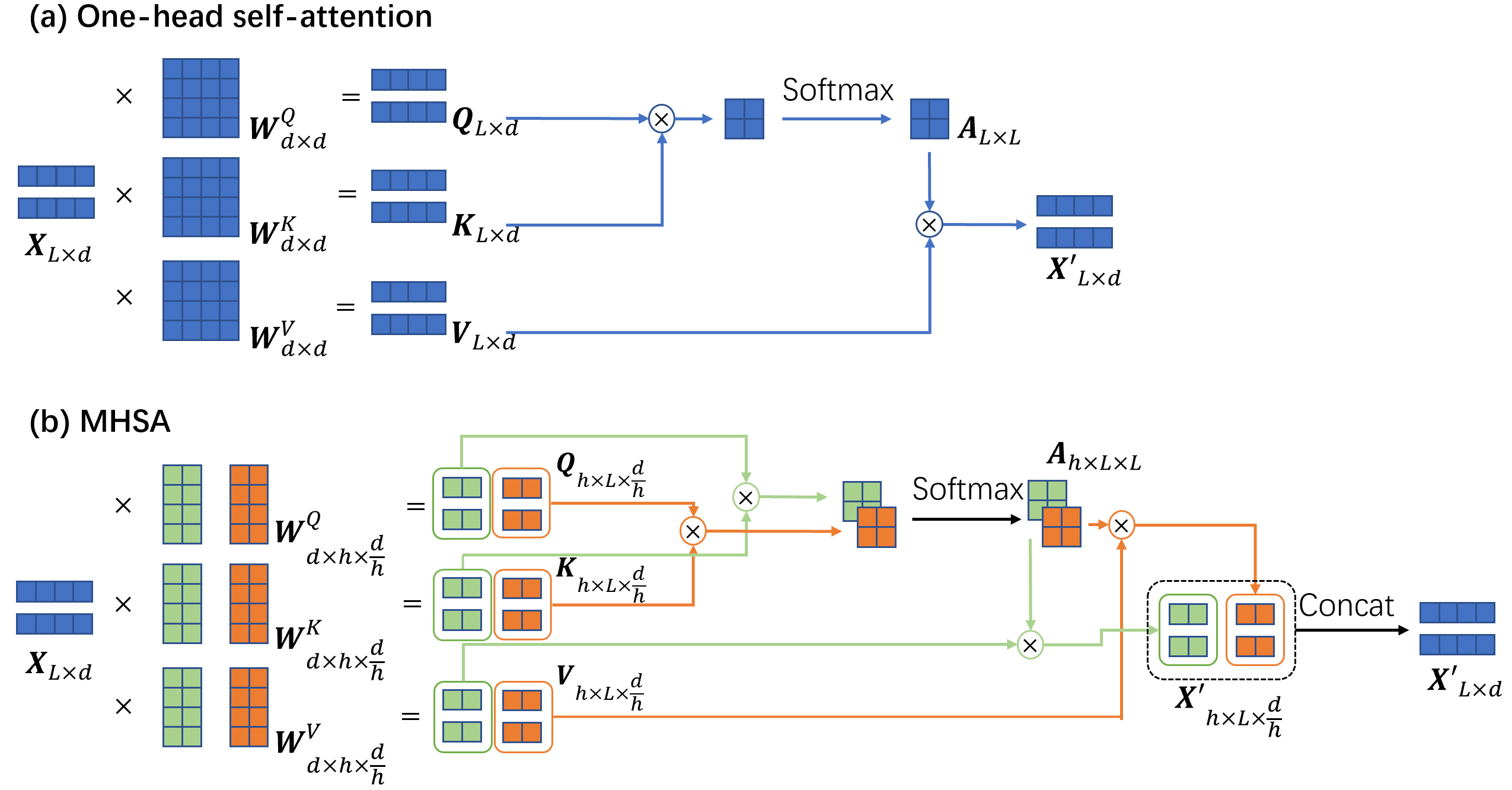}
    \caption{A schematic illustration of the self-attention mechanism within Transformer models and its capability for extracting long-range information. The top panel depicts a basic single-head self-attention mechanism. In this setup, the input is duplicated into three copies, each subjected to distinct linear transformations defined by the learnable matrices \(\boldsymbol{W}^Q\), \(\boldsymbol{W}^K\), and \(\boldsymbol{W}^V\). Two of these copies, \(\boldsymbol{Q}\) and \(\boldsymbol{K}\), are used for an inner dot product to compute similarities between different timestamps. These computed attention values are subsequently merged with the remaining \(\boldsymbol{V}\) copy to produce a new representation of the input sequence. The bottom panel showcases an extension to multi-head self-attention (MHSA). While retaining the same number of learnable parameters, MHSA partitions the linear transformations into separate blocks. The cross-matching between these blocks enables the capture of diverse correlations within sequences, accommodating different types of relevancy, such as varying time scales.}
    \label{fig:mhsa}
\end{figure*}

To elaborate, a schematic representation of this process is depicted in Fig.~\ref{fig:mhsa} (a), and can be summarized as follows:

\begin{itemize}
\item[1.] Self-attention starts by duplicating the input sequence and linearly projecting each copy into one of three forms: query, key, and value. These projections are steered by learnable parameters.

\item[2.] Each projection serves a unique purpose: the query seeks relevant information within the sequence, and the key aims to align with this query. This alignment is achieved through a dot product operation, measuring the correlation between various keys and queries. Pairs with high dot products are deemed strongly correlated and are thus allocated ``attention." Importantly, the dot product spans the entire sequence, allowing for attention to be distributed between any two timestamps, regardless of their separation.

\item[3.] The value vector then leverages these attention weights to decide the proportion of the original sequence's information to include in the output. As a result, the output vector, serving as the new sequence representation, is a weighted compilation based on these attention mechanisms.
\end{itemize}

To operationalize this, we consider an input sequence \(\boldsymbol{X} \in \mathbb{R}^{L \times d}\), where \(L\) represents the sequence length and \(d\) denotes the dimension of each vector. For notational convenience, we'll use the subscript \(L \times d\) to specify the dimensions for all tensors henceforth. Self-attention initiates the process by linearly projecting the input \(\boldsymbol{X}_{L \times d}\) into three separate forms: query \(\boldsymbol{Q}_{L \times d}\), key \(\boldsymbol{K}_{L \times d}\), and value \(\boldsymbol{V}_{L \times d}\). Mathematically, these projections are defined as:
\begin{eqnarray}
\boldsymbol{Q}_{L \times d} = \boldsymbol{X}_{L \times d}\boldsymbol{W^Q}_{d \times d}, \\
\boldsymbol{K}_{L \times d} = \boldsymbol{X}_{L \times d}\boldsymbol{W^K}_{d \times d}, \\
\boldsymbol{V}_{L \times d} = \boldsymbol{X}_{L \times d}\boldsymbol{W^V}_{d \times d},
\end{eqnarray}

\noindent
where \(\boldsymbol{W^Q}_{d \times d}, \boldsymbol{W^K}_{d \times d}, \boldsymbol{W^V}_{d \times d}\) are learnable matrices.

The next step involves calculating the similarity between the query and key vectors by taking an inner dot product over the \(d\) dimension, which yields an \(L \times L\) similarity matrix $(\alpha_{ij})$. To endow these similarities with a probabilistic interpretation, a softmax function is applied to responses belonging to each query to ensure the attention weights are both positive and normalized:
\[
\alpha_{ij} \mapsto \frac{e^{\alpha_{ij}}}{\sum_{k=1}^{L} e^{\alpha_{ik}}}.
\]
We call this normalized similarity matrix \(\boldsymbol{A}_{L\times L} = (A_{ij})\) attention map, where each entry \(A_{ij}\) is the normalized dot product between the \(i\)-th query vector \(\boldsymbol{q}^i_d\) and the \(j\)-th key vector \(\boldsymbol{k}^j_d\). This key feature sets self-attention apart from convolutional layers by measuring similarity across all timestamps.

Finally, a new representation for \(\boldsymbol{X}\) is generated by matrix-multiplying the attention matrix \(A\) with the value matrix \(\boldsymbol{V}\). This operation effectively weights a linearly transformed version of the original input data using the computed attention. Formally, for each self-attention layer, the input \(\boldsymbol{X}\) undergoes the following transformation:
\[
\boldsymbol{X} \mapsto {\rm softmax}(\boldsymbol{Q}\boldsymbol{K}^T)\boldsymbol{V}.
\]

\noindent
The resulting vectors extract specialized information from the entire sequence, taking into account which pairs of timestamps should be allocated additional ``attention."

In practice, it has been found \citep{attentionisallyouneed} that normalizing with the attention also with the feature dimension $\sqrt d$ facilitates the training of the neural networks. Therefore, self-attention, including in this study, is implemented as
\begin{equation}
    \boldsymbol{X} \mapsto {\rm softmax}\bigg(\frac{\boldsymbol{Q}\boldsymbol{K}^T}{\sqrt{d}}\bigg)\boldsymbol{V}.
\end{equation}

This straightforward approach, which involves linear transformations and computing attention via dot products, is termed `single-head' self-attention. While efficient, this model is somewhat limited in its expressiveness due to its reliance on a single set of weights. To enhance its versatility, the concept is extended to Multi-Head Self-Attention (MHSA), enabling the model to capture various types of correlations within the sequence.

In MHSA, \(h\) separate self-attention operations are performed, each with its own uniquely parameterized \(\boldsymbol{W^Q}, \boldsymbol{W^K}, \boldsymbol{W^V}\). Here, \(h\) denotes the number of heads in the model. To elucidate this, Fig.~\ref{fig:mhsa} illustrates how a two-head self-attention mechanism processes an input \(\boldsymbol{X}_{2 \times 4}\). Rather than employing a single set of weights, the weight tensors \(\boldsymbol{W^Q}_{4 \times 4}, \boldsymbol{W^K}_{4 \times 4}, \boldsymbol{W^V}_{4 \times 4}\) are partitioned into \(h=2\) segments. The queries, keys, and values from these different segments are then correlated independently to construct a range of attention-weighted arrays \(\boldsymbol{X}'\). When compared to single-head self-attention with an equivalent number of parameters, MHSA proves more adept at capturing diverse patterns, e.g. different time scales, within the sequence. 

\subsection{\textit{Astroconformer} Architecture}
\label{sec:Astroconformer}

The self-attention mechanism outlined earlier serves as the foundation of our model, \textit{Astroconformer}. However, to fully leverage the richness of stellar light curves, \textit{Astroconformer} integrates both convolutional layers and MHSA layers. The convolutional layers are adept at capturing local, short-range information, while the MHSA layers excel at extracting long-range correlations. In the sections that follow, we will briefly discuss the architecture of \textit{Astroconformer} and its essential components. Detailed explanations of the training techniques specific to \textit{Astroconformer} are deferred to Appendix~\ref{sec:train}.

Our primary focus is on individual quarters of \textit{Kepler} data, which serve as the input for \textit{Astroconformer}. Each quarter of \textit{Kepler} data spans approximately 90 days and features long-cadence data collected at intervals of 29.4 minutes. After the data preprocessing, as described in Section~\ref{sec:data}, the light curves are resampled at 30-minute intervals. Consequently, a 90-day light curve consists of 4320 timestamps. During the training process, we randomly crop segments containing 4,000 timestamps from each quarter's light curve. This strategy allows for parallel computation within each batch and serves a dual purpose as a form of data augmentation. By increasing the diversity of the input data, random cropping helps mitigate challenges associated with data scarcity.

\begin{figure}
    \centering
    \includegraphics[width=0.5\textwidth]{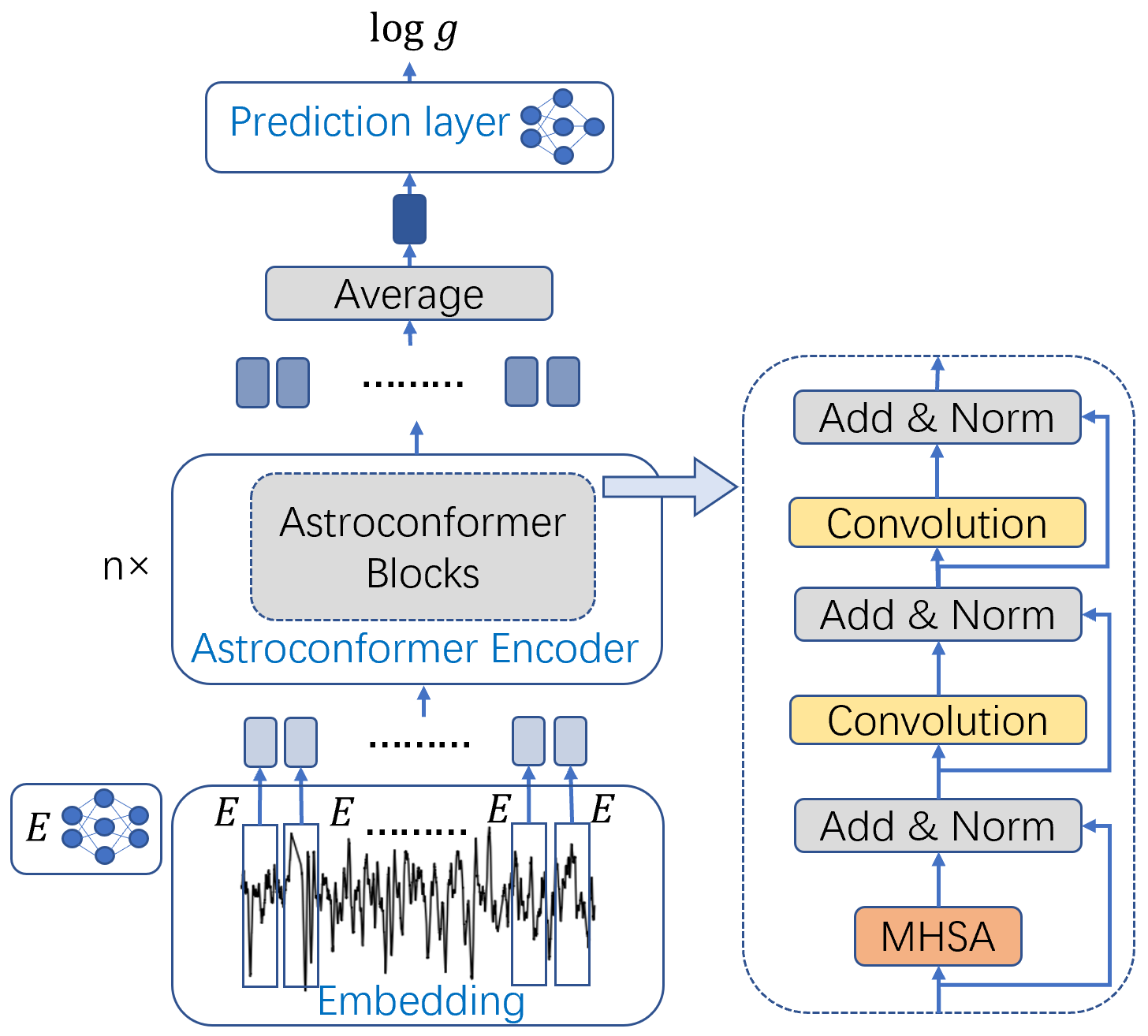}
    \caption{Architecture of \textit{Astroconformer}. The icon labeled \( E \) represents the MLP employed in the model. Stellar light curves are partitioned into patches of size 20 (corresponding to a time span of 10 hours), each of which is then transformed into a vector via a fully connected layer denoted by \( E\). These vectors are processed through the \textit{Astroconformer} encoder to extract local and global features. The \textit{Astroconformer} architecture consists of multiple such blocks, each incorporating a learnable MHSA module followed by two convolutional modules. To facilitate training, skip connections and subsequent layer normalization are applied between every pair of adjacent modules within each \textit{Astroconformer} block. Vectors output by the \textit{Astroconformer} encoder are aggregated through average pooling to produce the final representation of the entire light curve. This representation vector is subsequently fed into a final MLP for predicting the \(\logg\) of stars.}
    \label{fig:Astroconformer}
\end{figure}

\subsubsection{Embedding}
\label{sec:convembed}

Inspired by the Vision Transformer \citep{vit}, which divides an image into fixed-size patches and employs a Multi-Layer Perceptron (MLP) to project each into a high-dimensional space, \textit{Astroconformer} adopts a similar approach for transforming stellar light curves into sequences of vectors. As depicted in Fig.~\ref{fig:Astroconformer}, the input \textit{Kepler} data first passes through a patch-embedding layer, where the light curve is segmented into 200 patches, each comprising 20 timestamps. These patches are then converted into 128-dimensional vectors using an MLP denoted as \( E \). This transformation restructures the original one-dimensional input vector \(\mathbb{R}^{4000}\) into a sequence of vectors with dimensions \(\mathbb{R}^{200 \times 128}\). Here, the first dimension of 200 can be interpreted as a coarse temporal axis, while the second 128-dimensional axis represents the localized features captured by the MLP \( E \). When dealing with light curves that differ in length from the standard 4,000 timestamps, Astroconformer consistently follows the approach of segmenting these curves into patches of 20 timestamps. For instance, with Kepler Q1 data (~33 days, 1,600 timestamps), the curve is divided into 80 patches, each individually embedded into a 128-dimensional space. No padding values at the end are introduced to the input.

This embedding serves dual objectives. Firstly, the MLP \( E \) focuses on extracting localized, short-range features, covering spans of 20 timestamps or approximately 10 hours. Secondly, by mapping the input into a 128-dimensional feature space, the model is well-positioned to exploit the power of MHSA, enabling it to cross-correlate different features across various time spans.

\subsubsection{\textit{Astroconformer} Encoder}
\label{sec:encoder}

The (200, 128)-dimensional vector sequence, produced by the embedding layer, is subsequently inputted into the \textit{Astroconformer} encoder. This encoder is designed to simultaneously capture both global long-range and localized short-range information from the sequence of vectors.

In our architecture, the \textit{Astroconformer} encoder is made up of multiple \textit{Astroconformer} blocks. Each of these blocks comprises an 8-head MHSA layer and two convolutional modules. Each convolutional module includes a convolutional layer, batch normalization \citep{batchnorm}, and the Swish activation function \citep{swish}. The settings for each convolutional layer are a kernel size of 3, a stride of 1, and same padding. The number of channels in each embedding vector remains constant at 128 throughout the entire forward process of the encoder. Our results indicate that the inclusion of convolutional layers following each MHSA layer enhances performance. This is possibly because MHSA modules, while capable of capturing long-range correlations, lack inherent assumptions about data locality, which often necessitates larger datasets for optimal learning compared to CNNs. In our specific scenario, the relatively limited dataset ($\sim$14,000 stars) could have made it difficult for a model solely reliant on MHSA layers to converge satisfactorily, motivating our incorporation of convolutional layers.

Compared to the original Conformer architecture \citep{conformer}, we have omitted the feed-forward modules within the conformer block, as our experiments show that they add a significant number of parameters without substantially improving performance. Focusing the primary parameters on modules that enable meaningful interactions between patches appears to be a more parameter-efficient strategy. Additionally, we refrained from employing sparse convolutions like depthwise and pointwise convolutions to ensure a fair comparison with ResNet-18, as discussed in Section~\ref{sec:vscnn}. 

It is worth noting that MHSA is inherently agnostic to the sequence order of its input. To address this limitation and make sequence order information available to the model, we employ Rotary Positional Encoding (RoPE) \citep{RoPE} in each MHSA module. Briefly, RoPE compensates for the MHSA's lack of sensitivity to input order by multiplying the query and key vectors with complex phases that are proportional to their positions, $\boldsymbol{q}_m' = \boldsymbol{q}_me^{i\omega m}, \boldsymbol{k}_n' = \boldsymbol{k}_ne^{i\omega n}$, where $\omega$ is a fixed frequency parameter. This ensures that their relative positions are encoded into the resulting inner products $\langle\boldsymbol{q}_m', \boldsymbol{k}_n'\rangle = \boldsymbol{q}_m\cdot\boldsymbol{k}_ne^{i\omega (m-n)}$\footnote{We refer to the original paper for the details, including the choice of $\omega$.}.

\subsubsection{Pooling and Prediction Layer}
\label{sec:pooling}

The Embedding layer and the \textit{Astroconformer} encoder together generate a final patch embedding sequence with dimensions $200 \times 128$ derived from the input light curve. We apply an average pooling to patch embeddings to the temporal dimension and obtain a final 128-dimensional embedding of the light curve (Fig.~\ref{fig:Astroconformer}). Since we are dealing with individual patches of the light curves and aggregating all the information through average pooling, this enables \textit{Astroconformer} to cope with light curves of variable lengths.

Subsequently, this 128-dimensional final embedding is processed through a two-layer MLP with a dropout probability of 0.3. This serves as the final layer tasked with predicting $\logg$.

%
%
%
%
%
%
\section{Data}
\label{sec:data}

\subsection{Sample Selection}

\begin{figure}
    \centering
    \includegraphics[width=0.5\textwidth]{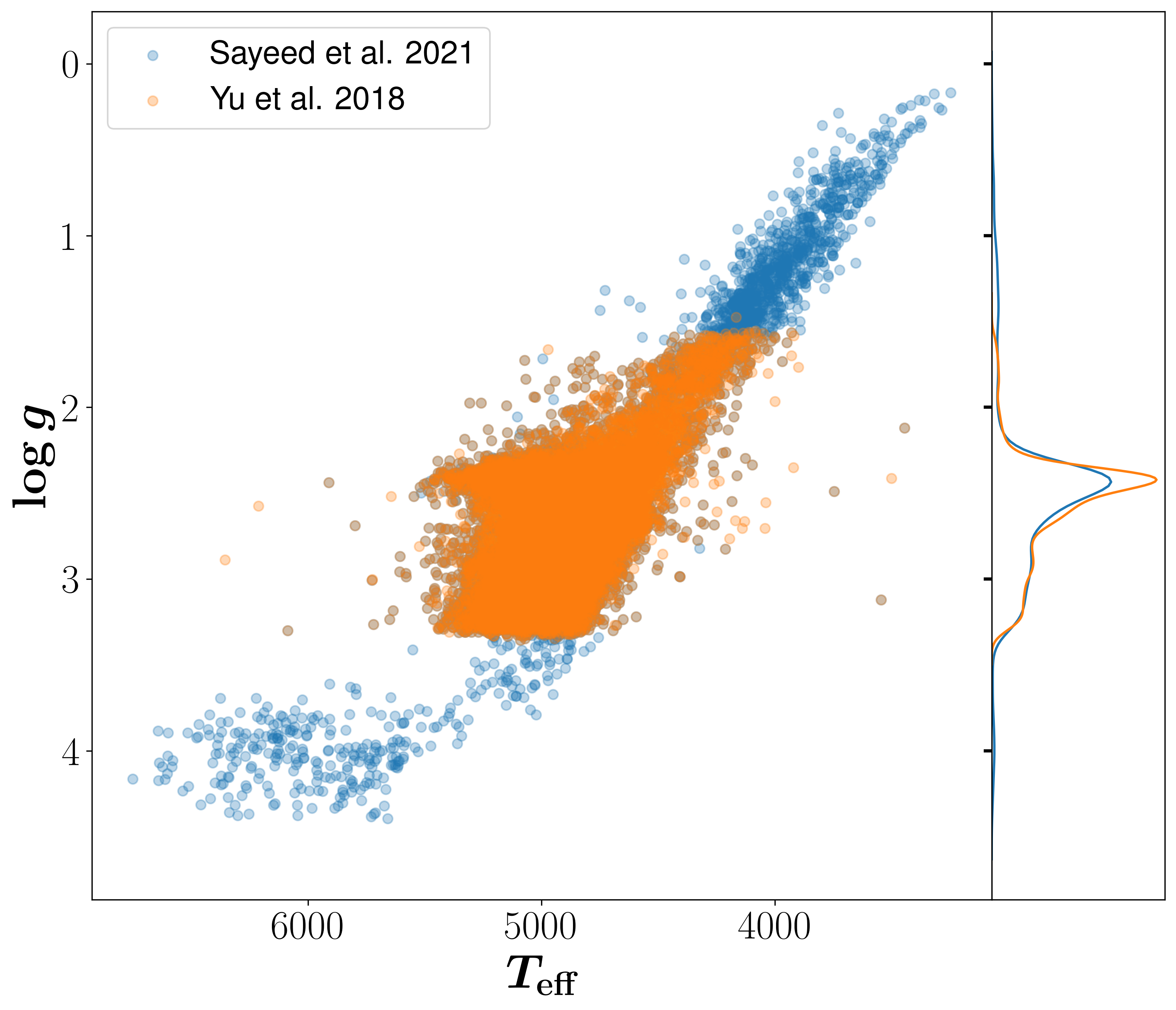}
    \caption{Distribution of surface gravity versus effective temperature for our two data sets. The data set from \citet{yu18} consists of 16,094 giant stars with $\logg$ values ranging from 1.5 to 3.3. Based on \citet{yu18}, the data set from \citet{swan} further includes turnoff stars and extends to the tip of the red giant branch (TRGB), while a data curation detailed in text is executed, leading to 14,003 stars in the data set.}
    \label{fig:data}
\end{figure}

In this study, we leverage two data sets, \citet{swan}, and \citet{yu18} to facilitate a direct comparison between \textit{Astroconformer} and other methods (Fig.~\ref{fig:data}). 

\citet{yu18} systematically characterized solar-like oscillations and granulation for 16,094 oscillating red giants, using end-of-mission long-cadence data. These stars feature $\logg$ values between 1.5 and 3.3. The typical uncertainties of their $\nu_{\max}$ and asteroseismic $\logg$ estimations are 1.6\% and 0.01-0.02 dex, respectively. For the input data, we include all available quarters of light curves for each star, provided that the quarter contains at least 3,000 valid observations (quality flag = 0). This criterion results in a data set comprising 147,731 quarters of light curves for 15,874 red giants. This high-quality data is used for our training data as well as the comparison done in Section~\ref{sec:vsastero}.

To further compare with the state-of-the-art machine learning method in asteroseismology, we also include an analysis with \citet{swan}.
\citet{swan} cross referenced two asteroseismic surface gravity catalogue from \citet{mathur} and \citet{yu18}. \citet{mathur} compiled asteroseismic stellar properties from 11 catalogues (see references therein). Stars supplemented by \citet{mathur} include high-luminosity red giants with $\logg < 2$, along with subgiants and main-sequence stars. In Section~\ref{sec:vsswan} and Section~\ref{sec:vscnn}, the same asteroseismic $\logg$ as \citet{swan} from these two work are used for training and evaluation.

Furthermore, to ensure a fair comparison with {\sc the swan}, we utilize an identical dataset of single-quarter long-cadence light curves as our input data. {\sc the swan} adopted specific data curation schemes. It excludes light curves of rotating stars, classical pulsators, eclipsing binaries, and \textit{Kepler} exoplanet host stars. Additionally, light curves missing observations from \textit{Kepler} Q7-11, or lacking stellar parameter estimates from \citet{berger}, are omitted as stellar radius data is essential for their subsequent light curve smoothing.  These criteria yield a final dataset of 14,003 stars, boasting reliable 0.2--4.4 $\logg$ estimates across a range of evolutionary stages, including turnoff stars, red-giant-branch (RGB) stars, and red clump stars. 

\subsection{Light Curve Pre-Processing}

For the preprocessing of these light curves, we adhere to the following steps:

\begin{itemize}
\item[1.] Only flux values with good quality flags (flags = 0) are retained. We apply sigma-clipping within a rolling window of 50 timestamps, clipping 3$\sigma$ outliers.
  
\item[2.] We employ a Savitzky-Golay filter to eliminate any low-frequency variations \citep{astero_preprocess}. The filter size, denoted as $d$, is chosen to vary with the stellar radius according to the formula $d[\mu \mathrm{Hz}] = 0.92 + 23.03 e^{-0.27 \mathrm{R} / \mathrm{R}_{\odot}}$ \citep{swan}. This ensures that the filter size increases with $\nu_{\max}$, allowing for the adaptive removal of long-period variations.

\item[3.] To address any missing values that may have been introduced in Step 1, we use linear interpolation to fill in the gaps. The interpolated segments are shorter than 2.1 days for 90\% stars, with the largest gap being 13.6 days.

\item[4.] To further aid in model training, which benefits from inputs having similar orders of magnitude in terms of standard deviation, we normalize the light curve using the following equation:
\begin{equation}
    \frac{\boldsymbol{x}}{\rm{std}(\boldsymbol{x})} \times \frac{1}{\rm{ln}(\rm{maxstd}/\rm{std}(\boldsymbol{x}))+1} \mapsto \boldsymbol{x},
\end{equation}
where \(\rm{maxstd}\) represents the largest standard deviation observed across all light curves. This normalization ensures that after preprocessing, standard deviations of light curves are restricted between 0.1-1.
\end{itemize}

It is important to note that the normalization process serves another purpose: to preserve the relative amplitude ranking of stellar oscillations. The amplitude is a critical attribute of these oscillations, and this normalization method ensures that light curves with inherently larger oscillations amplitudes maintain their amplitude rankings even after processing.

%
%
%
%
%
%
\section{Results}
\label{sec:result}

In this section, we discuss the advantages of \textit{Astroconformer} in comparison with other approaches. Section~\ref{sec:vsswan} and Section~\ref{sec:vscnn} contrast \textit{Astroconformer} with other proposed machine learning methods like {\sc the swan} and CNNs, underscoring the utility of Transformer models in optimally leveraging information from stellar light curves. Following that, Section~\ref{sec:vsastero} explores a comparative analysis between \textit{Astroconformer} and two asteroseismic pipelines. Our results suggest that \textit{Astroconformer} could offer a precise and reliable method for predicting stellar parameters, particularly when data is constrained in timespan—an issue of critical importance for data from missions like K2, TESS.

\subsection{Compared with k-NN Based Methods}
\label{sec:vsswan}

We first compare the performance of \textit{Astroconformer} with that of {\sc the swan}. We focus on the same dataset as used in {\sc the swan}, which comprises 14,003 stars with asteroseismic $\logg$ values ranging from 0.2 to 4.4 dex. Our analysis is also confined to \textit{Kepler} data with a 90-day duration to enable a direct comparison. For \textit{Astroconformer}, the 14,003 light curves are divided into training, validation, and test sets in proportions of 72\%, 8\%, and 20\%, respectively. A 5-layer \textit{Astroconformer} is trained on the training set, and the best-performing model is selected based on its performance on the validation set. This model is then used to generate predictions on the test set. We emphasise that our training labels are from the \citet{mathur} and \citet{yu18}, and we are only evaluating the test predictions by {\sc the swan}.

To facilitate a direct comparison, we also evaluate {\sc the swan}'s test prediction on the same test set. Note that \textit{Astroconformer} operates on a smaller training set compared to {\sc the swan}. In its approach to inferring surface gravity, {\sc the swan} utilizes a linear regression method based on a star's nearest neighbors, effectively implementing leave-one-out cross-validation. Conversely, \textit{Astroconformer} sets aside 20\% of the data as a test set.

As for performance metrics, we opt for the RMSE between the predictions and the asteroseismic $\logg$ values instead of the median absolute deviation $\sigma_{\text{mad}}$ employed by {\sc the swan}. This choice is motivated by our observation that $\sigma_{\text{mad}}$—which is less sensitive to outliers due to the use of the median—may not fully capture the nuances of the dataset.

\begin{figure}
    \centering
    \includegraphics[width=0.42\textwidth]{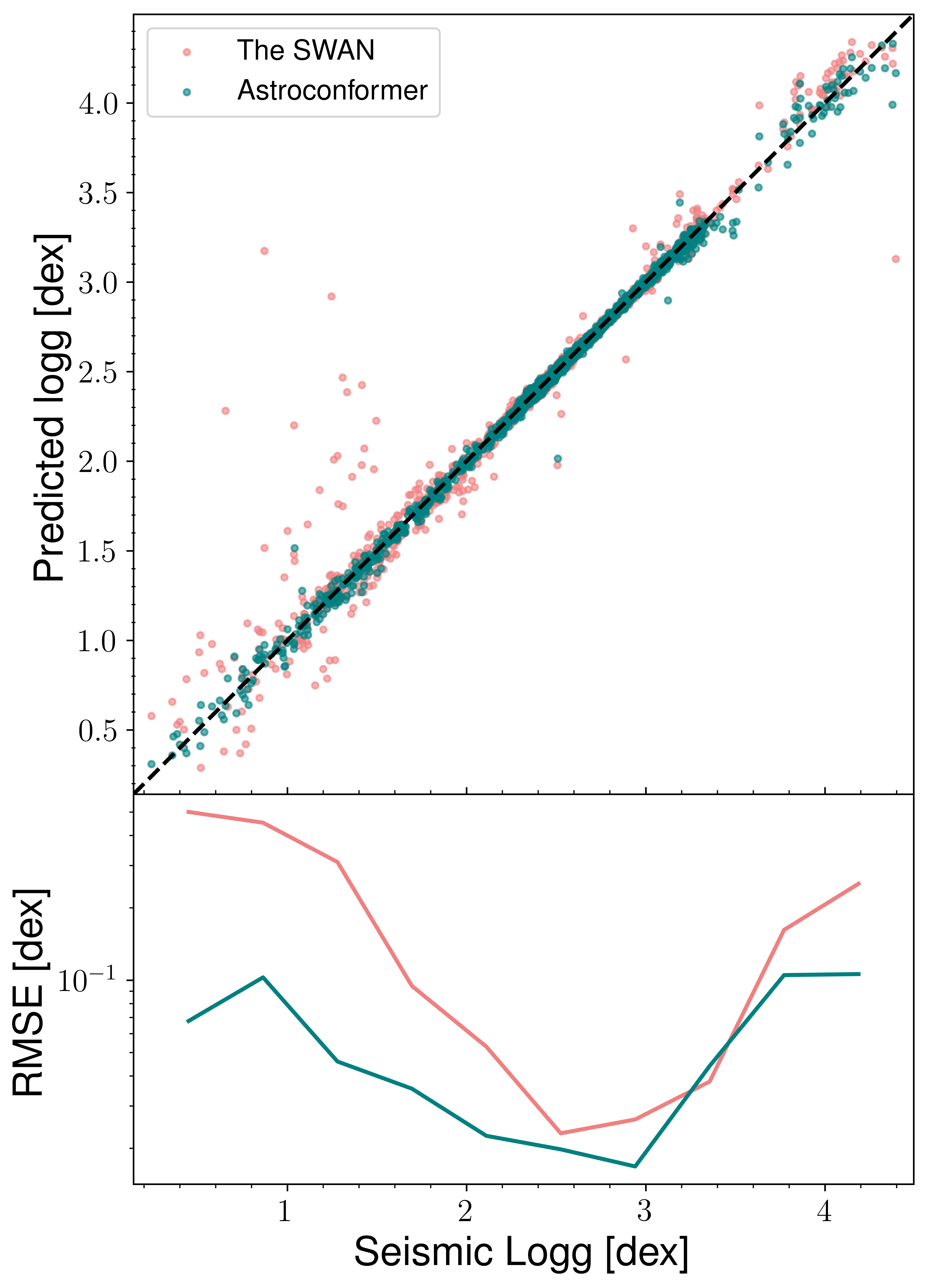}
    \caption{A comparative study of \textit{Astroconformer} and the k-nearest neighbor-based method employed in {\sc the swan}. \textit{Astroconformer} demonstrates enhanced generalization and yields fewer outlier inferences, particularly in the upper giant branch. It also consistently outperforms {\sc the swan} in terms of RMSE across the full $\logg$ range. The top and middle panels show the inferred $\logg$ values obtained from \textit{Astroconformer} and {\sc the swan}, respectively, plotted against the asteroseismic $\logg$. The bottom panel illustrates the running mean in RMSE (with a window size of 0.4 dex) for both algorithms.}
    \label{fig:vsswan}
\end{figure}

Fig.~\ref{fig:vsswan} illustrates the performance comparison between \textit{Astroconformer} and {\sc the swan}. {\sc the swan}'s RMSE hovers around 0.03 dex at $\logg = 3$, and then increases to between 0.3 and 0.5 dex at both the subgiant phase and the upper part of the red giant branch. In contrast, \textit{Astroconformer} consistently outperforms {\sc the swan} across the entire $\logg$ spectrum, achieving a root-mean-square error (RMSE) of at most 0.1 dex in the 0.2--4.4 $\logg$ range. It performs particularly well around $\logg = 3$, where the training sample is denser, with a minimum RMSE of 0.017 dex. 

The result is perhaps unsurprising, k-NN algorithms often falter in high-dimensional feature spaces and are highly sensitive to sampling density.  This is evident at the upper giant branch, where the training sample is sparse, leading {\sc the swan} to incur more outliers and reduced robustness. In contrast, deep learning methods like \textit{Astroconformer} aim to generalize by learning underlying features, making them more robust even when training samples are sparse.

It is noteworthy that \textit{Astroconformer} successfully determines \(\logg\) to an RMSE $< 0.1$ dex for dwarf stars with \(\logg > 4\) and high-luminosity red giants with \(\logg<1\). Considering that oscillation signals in dwarf stars are too brief for detection in Kepler long cadence data, this achievement indirectly demonstrates \textit{Astroconformer}'s capability to utilize granulation information within the dataset. Similarly, for giants near the tip of the red giant branch, their extended oscillation periods $>10$ days can not be resolved well by the typical 90-day span. \textit{Astroconformer}'s ability to accurately infer \(\logg\) for these stars further indicates its proficiency in leveraging information beyond just the oscillation peak.

Finally, we note that a certain degree of scatter may arise from uncertainties in the training labels. Given that the asteroseismic $\logg$ values have an uncertainty of 0.01--0.02 dex, this factor could contribute to the error. However, considering that this uncertainty is at most half of the prediction errors observed in both algorithms, we anticipate its impact on our overall analysis to be minimal. For a discussion on addressing label uncertainty, please refer to Section~\ref{sec:limitation}.

\subsection{Transformer Models versus Convolutional Neural Networks}
\label{sec:vscnn}

We now turn our attention to comparing \textit{Astroconformer} with various time-domain deep learning models. As mentioned in the introduction, this study is motivated by the hypothesis that subtle yet crucial information in stellar light curves may be concealed at longer time scales.

\begin{figure}
    \centering
    \includegraphics[width=0.5\textwidth]{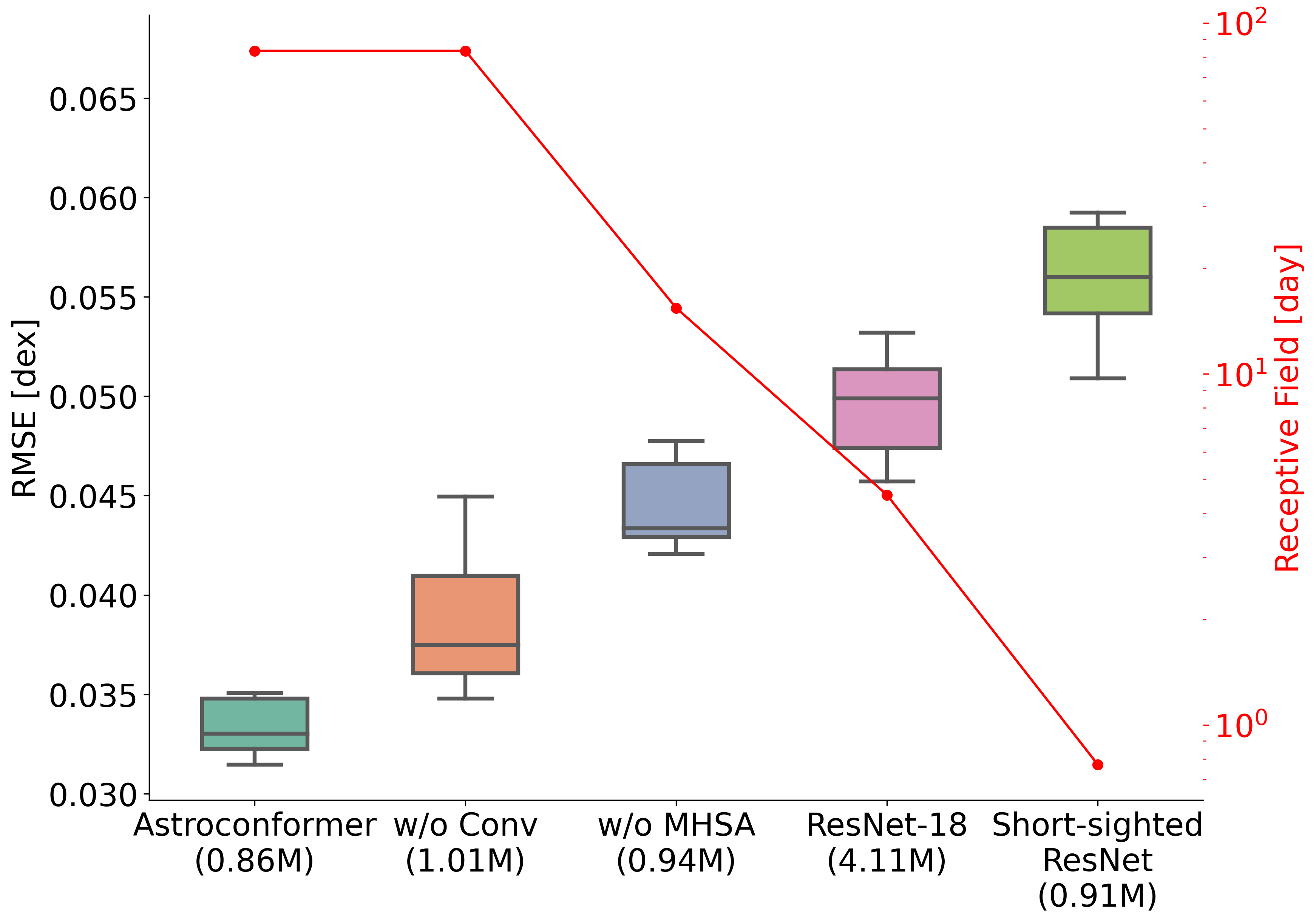}
    \caption{Ablation study illustrating the RMSE for \(\logg\) inference across various models with different receptive fields. The respective numbers of parameters for each model are annotated alongside their names. The boxplot depicts the distribution of RMSE values on the test set, as assessed by 10-fold cross-validation for each model. We omit outliers more than 1.5 times the interquartile range (IQR) above the third quartile to investigate the general trend. Receptive fields for the models are indicated by red lines on a logarithmic scale. The figure reveals that as the receptive field size diminishes, the median RMSE for each model correspondingly increases. Moreover, Transformer-based models outperform a conventional CNN, ResNet-18, despite the latter having four times more parameters. Furthermore, the first two models on the plot indicate that while MHSA is crucial for performance, the incorporation of convolutional modules further enhances \textit{Astroconformer}'s effectiveness.}
    \label{fig:vscnn}
\end{figure}

In our ablation study, we study the performance of five distinct models, each characterized by varying extents of receptive fields: (a) \textit{Astroconformer}; (b) a variant of \textit{Astroconformer} employing only MHSA; (c) \textit{Astroconformer} devoid of MHSA; (d) the ResNet-18 model; and (e) a modified ResNet architecture, which we designate as `Short-Sighted ResNet' (Fig.~\ref{fig:vscnn}).

To ensure a rigorous comparison, each model under consideration adopts an architecture similar to Astroconformer. This architecture incorporates identical components such as MHSA, convolutional modules, an average pooling layer, and a prediction layer, as delineated in Section~\ref{sec:Astroconformer}. For each module, we either substitute the MHSA with convolutional modules or vice versa. Where appropriate, we also adjust the number of blocks to maintain roughly equivalent counts of learnable parameters across the neural networks. An exception to this is ResNet-18, which remains consistent with its established architecture as cited in \citet{resnet}. The specifics are itemized in Table~\ref{tab:modeldetail}.

As illustrated in Table~\ref{tab:modeldetail}, the `without convolution' variant (b) replaces all convolutional modules with MHSA modules while retaining the same global receptive field inherent to \textit{Astroconformer}. The `without MHSA' variant (c) substitutes all MHSA modules with convolutional modules, yielding a receptive field of 15 days. Variant (e) `Short-Sighted ResNet,' serves as an extended modification of the without MHSA variant (c), but omits the embedding layer. By accepting raw timestamps instead of patch embeddings as inputs, its receptive field is truncated to 18 hours. Lastly, for benchmarking purposes, we include ResNet-18, known for its proficiency in time series \citep{tsc}, featuring a 4.5-day receptive field.

\begin{table}
	\centering
	\caption{The five models conducted in the ablation study, and their constituent submodules within each block, the total number of blocks, and the utilization of an embedding layer.}
	\label{tab:modeldetail}
	\begin{tabular}{lccc} 
	\toprule
Model         & Block& \# Blocks& Embedding\\ 
\midrule
\textit{Astroconformer} & MHSA, Conv, Conv& 5& \cmark\\
 without Conv& MHSA, MHSA, MHSA& 5&\cmark\\
 without MHSA& Conv, Conv, Conv& 6&\cmark\\
 Short-sighted ResNet& Conv, Conv, Conv& 6&\xmark\\
 ResNet-18& -& -&\xmark\\
		\hline
	\end{tabular}
\end{table}

We subject all models to 10-fold cross-validation, using the same dataset outlined in Section~\ref{sec:vsswan}, which includes 14,003 high-quality \textit{Kepler} samples with \(\logg\) values ranging from 0.2 to 4.4. Instead of investigating running RMSE as in Section~\ref{sec:vsswan}, here we use overall RMSE as the performance measure of different models. This is motivated by the fact that long-range correlation such as granulation is important for $\logg$ estimation for all stars, and a smaller overall RMSE is expected if a model can extract long-range correlation better. The overall RMSE values from the 10 cross-validation tests are presented in Fig.~\ref{fig:vscnn}, while we omit outliers more than 1.5 times the interquartile range (IQR) above the third quartile to investigate the general trend. In this figure, the boxplot represents the distribution of RMSE values across the different cross-validation runs for each model, while the red solid line indicates the corresponding receptive field for each model. We report the median RMSE of the five models to be 0.033 dex, 0.037 dex, 0.043 dex, 0.050 dex, and 0.056 dex, respectively.

\begin{figure*}
    \centering
    \includegraphics[width=1.0\textwidth]{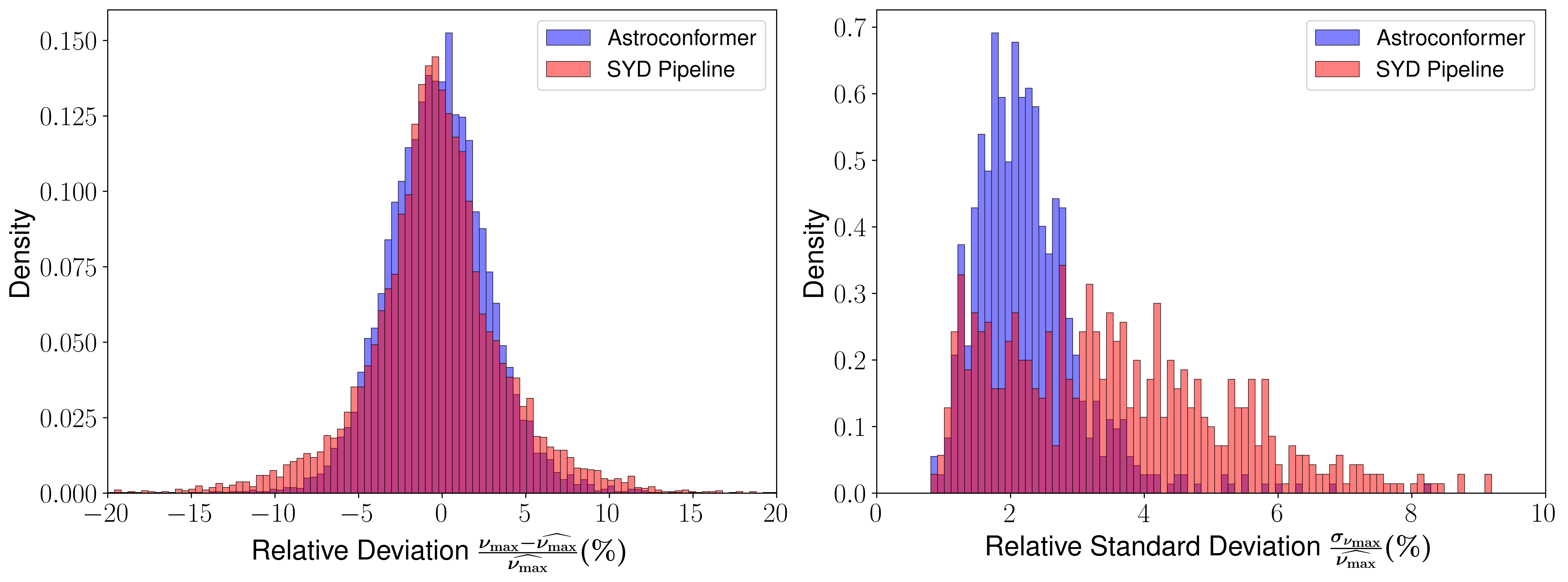}
    \caption{We compare the predictive accuracy and consistency of \textit{Astroconformer} and the SYD pipeline using single-quarter \textit{Kepler} light curves. \textit{Left Panel:} The distribution of relative deviation for \(\nu_{\max}\) predictions from \textit{Astroconformer} and the SYD pipeline is shown, each based on single-quarter light curves and compared to the `ground truth' \(\widehat{\nu_{\max}}\) provided by \citet{yu18} using full four-year \textit{Kepler} data. \textit{Astroconformer} achieves a relative median absolute error of 1.93\%, compared to the SYD pipeline's result of 2.10\%. \textit{Right Panel:} The relative standard deviations in predictions for each star across various quarters are displayed. \textit{Astroconformer} has a median relative standard deviation of 2.10\%, compared to the SYD pipeline's median of 3.37\%.} 
    \label{fig:vsastero}
\end{figure*}

This figure illustrates the combined influence of both receptive field size and MHSA on the accuracy of \(\logg\) extraction from stellar light curves. A clear performance decline is observed as the receptive field size decreases across all five models. Importantly, this decline is not affected by the number of parameters, as demonstrated by ResNet-18. Despite having four to five times more parameters than the other models, ResNet-18 still adheres to the same trend. Additional evidence of the importance of receptive field size can be seen when comparing `\textit{Astroconformer} without MHSA' and `Short-sighted ResNet.' The key difference between these models lies in their input: the former processes patches of the light curve, thereby having a receptive field 20 times larger than the latter, which deals directly with light curve timestamps. This expanded receptive field contributes to a significant RMSE reduction from 0.056 dex to 0.043 dex .

Similarly, the only distinction between \textit{Astroconformer} and `\textit{Astroconformer} without MHSA' is the incorporation of the MHSA module in the former, which enhances its ability to assimilate global information. This enhancement results in a significant decrease in RMSE from 0.043 dex to 0.033 dex for \textit{Astroconformer}. These observations underscore the pivotal role played by receptive fields. In addition to local information like amplitudes that can be extracted by every model, long-range correlations resulting from granulation and oscillations is also important for precise \(\logg\) extraction from light curves.

Finally, a comparison between the default \textit{Astroconformer} model and the `\textit{Astroconformer} without Conv' model demonstrates the advantages of combining convolutional modules with self-attention mechanisms. Although the underlying Transformer architecture of \textit{Astroconformer} is theoretically capable of learning the convolution operation, the scarcity of labeled data makes the inclusion of convolutional modules particularly beneficial. Light curve data likely encompass both long-range and short-range correlations. By utilizing both convolutional modules and MHSA, \textit{Astroconformer} is better equipped to leverage these local and global correlations for more accurate predictions. This serves to underscore the significance of receptive field size, as well as the synergistic benefits of incorporating self-attention with convolution, in the modeling of long sequences of stellar light curves.

\subsection{Comparison with Asteroseismic Pipelines}
\label{sec:vsastero}
In addition to comparing \textit{Astroconformer} with other deep learning methods, this section evaluates its performance against traditional asteroseismic pipelines. Asteroseismic pipelines such as SYD typically focus on identifying \(\nu_{\max}\) by fitting and removing granulation and the white noise background in the Fourier spectrum and searching for the peak in the oscillations power excess. To facilitate a more rigorous comparison, we will study \(\nu_{\max}\) as the focal stellar parameter. This is driven by the understanding that asteroseismic \(\logg\) are derived from the scaling laws, which also take into account external information such as \(\teff\), which is not assumed in \textit{Astroconformer}.

The challenge of benchmarking \textit{Astroconformer} against asteroseismic pipelines arises from the absence of a priori ground truth labels for \(\nu_{\max}\). To overcome this hurdle, we utilize high-quality \(\nu_{\max}\) labels (with an uncertainty of approximately \(1.6\%\)) provided by \citet{yu18}, based on end-of-mission ($\sim$4 years) \textit{Kepler} data. 

\subsubsection{On 90-day Light Curves}
\label{sec:vssyd}
An 20-layer \textit{Astroconformer} is trained from scratch on the identical dataset in \citet{yu18} (refer to Appendix~\ref{sec:train} for training details). We then apply both \textit{Astroconformer} and the SYD pipeline to a test set containing 9,321 single-quarter light curves from \textit{Kepler} Quarters 2-16 except for Q8, 12\footnote{We omit Q8, 12due to an insufficient number of valid data points with a flag of 0, as discussed in Section~\ref{sec:data}.} (13 quarters in total) for 717 red giants\footnote{The seemingly limited test set size is because only 717 red giants have all 13 quarters of light curves out of the initial 1000 test stars.}. We utilize two key metrics for evaluation: relative deviation and relative standard deviation, aimed at assessing prediction accuracy and internal consistency, respectively. 

For relative deviation, we compute the fractional difference \( (\nu_{\max} - \widehat{\nu_{\max}})/\widehat{\nu_{\max}}\), comparing the predicted \(\nu_{\max}\) values obtained from single-quarter light curves against the `ground truth' \(\widehat{\nu_{\max}}\) values from \citet{yu18}. In terms of relative standard deviation, we calculate the sample standard deviation $\sigma_{\nu_{\max}}$ across all 13 predictions, each derived from different quarters' light curves for the same star, and then divide this by \(\widehat{\nu_{\max}}\). Essentially, the relative deviation evaluates how closely the predicted \(\nu_{\max}\) from single-quarter light curves approximates the `ground truth' from the long time-baseline derivation, while the relative standard deviation assesses the internal consistency of predictions across different quarters.

In Fig.~\ref{fig:vsastero}, we present the relative deviation on the left-hand side and the relative standard deviation on the right-hand side. While the relative deviations from both \textit{Astroconformer} and the SYD pipeline are broadly comparable, the SYD pipeline exhibits a longer tail in its distribution. Quantitatively, \textit{Astroconformer} achieves a relative median absolute error of 1.93\%, in comparison to the 2.10\% observed for the SYD pipeline. Moreover, \textit{Astroconformer}'s predictions are more consistent. The right panel of Fig.~\ref{fig:vsastero} demonstrates that \textit{Astroconformer}'s relative standard deviation is statistically much smaller than that of the SYD pipeline. Specifically, the relative median standard deviation for \textit{Astroconformer} is 2.13\%, in comparison to the 3.37\% observed for the SYD pipeline.

It is noteworthy that although \textit{Astroconformer} relies on training labels for \(\nu_{\max}\) obtained from the SYD pipeline's analysis of end-of-mission \textit{Kepler} light curves ($\sim$4 years), it provides predictions that are closer to `ground truth' $\widehat{\nu_{\max}}$ and more internally consistent than that of the SYD pipeline when applied to single-quarter light curves ($\sim$90 days). \textit{Astroconformer} leverages information beyond just the oscillations signal. As a result, even when working with single-quarter data, the model benefits from multiple constraints across information from the entire light curves, its corresponding power spectra but also the subtle correlation between the phases, enabling it to make precise and consistent predictions. 

It is a plausible consideration that oscillation modes frequently exhibit stochastic variations over time, as referenced in \citet{astero_lifetime}. Such variability could explain the SYD pipeline's deviations from global time-averaged quantities. Conversely, \textit{Astroconformer} is specifically trained to predict time-averaged \(\nu_{\max}\) as outlined in \citet{yu18}. This training approach accounts for its consistent performance across various quarters, indicating its capacity to harness additional information from the time series, extending beyond the mere oscillation modes.

\subsubsection{On 30-day Light Curves}
\label{sec:vshekker}

To more effectively illustrate the value of \textit{Astroconformer} as a supplement to existing pipelines, we delve into a scenario where light curves only have a 33-day timespan. We compare \textit{Astroconformer} with asteroseismic results in \citet{Hekker_2011} based on the same 33-day Q1 light curves. \citet{Hekker_2011} conducted an asteroseismic characterization for 16,511 red giants using their Q1 data, with oscillations detected in 10,956 (71\%) of these stars.

We fine-tune the 20-layer Astroconformer, as discussed in the previous section, on randomly cropped 30-day segments from the Q2-16 dataset. We then perform \(\nu_{\max}\) inference on Q1 light curves for 14,997 red giants, as defined in \citet{Hekker_2011}. Notably, all these samples possess more reliable \(\nu_{\max}\) labels from \citet{yu18}, derived from end-of-mission light curves that extend beyond these 30-day segments. Within this group, the pipeline from \citet{Hekker_2011} successfully detected oscillations in approximately 70\% of the stars (10,457 stars), marked as ``Detection". The remaining 30\% (4,540 stars) did not exhibit detectable oscillations and are consequently labeled as ``Non-detection".

\begin{figure}
    \centering
    \includegraphics[width=0.50\textwidth]{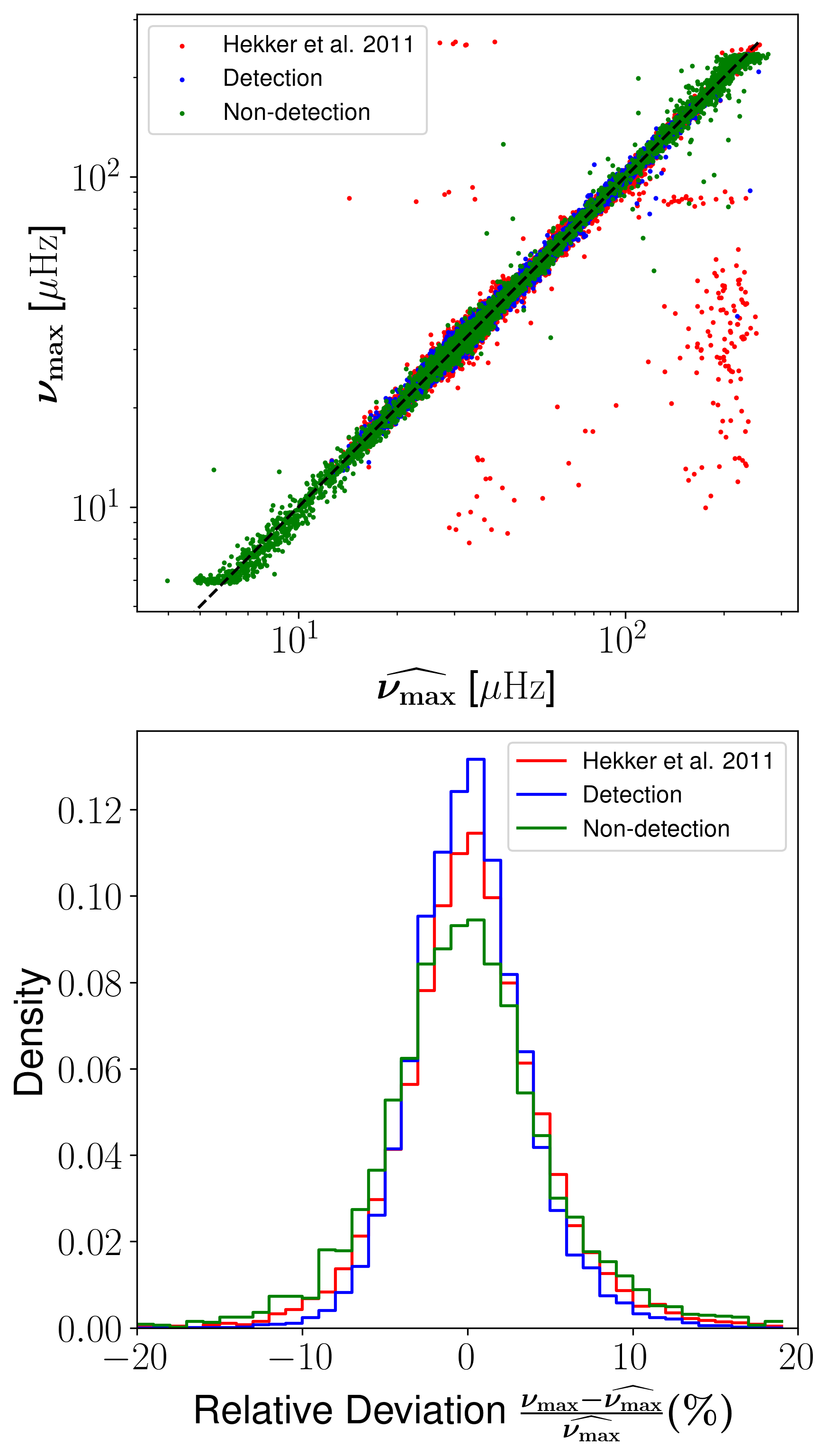}
    \caption{A comparative study of \textit{Astroconformer} and \citet{Hekker_2011} applied to Kepler Q1 light curves with an only 33-day time span. The top panel plot $\nu_{\max}$ estimations from both approaches against `ground truth' $\widehat{\nu_{\max}}$ from \citet{yu18}. Results of \textit{Astroconformer} are categorized into two subgroups, `Detection' and `Non-detection', depending on their detectability in \citet{Hekker_2011}. The bottom panel illustrates the relative deviation from $\widehat{\nu_{\max}}$. Astroconformer's relative median absolute error for the `Detection' group stands at 2.14\%, smaller than the 2.52\% of the conventional pipeline from \citet{Hekker_2011}. Even in the `Non-detection' subgroup where \citet{Hekker_2011} fails to identify oscillation modes, \textit{Astroconformer} maintains a relative median absolute error of 2.91\%.}
    \label{fig:vshekker}
\end{figure}

We assess the \(\nu_{\max}\) estimations made by \citet{Hekker_2011} and \textit{Astroconformer}, and juxtapose these with the `ground truth' \(\nu_{\max}\) from \citet{yu18}, denoted as \(\widehat {\nu_{\max}}\). In the top panel of Figure~\ref{fig:vshekker}, we display the \(\nu_{\max}\) estimations by \citet{Hekker_2011} and \textit{Astroconformer}, aligned against \(\widehat {\nu_{\max}}\). The \textit{Astroconformer} estimates are categorized into two subgroups, `Detection' and `Non-detection', based on the availability of $\nu_{\max}$ estimation in \citet{Hekker_2011}. Notably, the $\nu_{\max}$ estimations from \citet{Hekker_2011} apply exclusively to the `Detection' group. In the bottom panel, we present a histogram of the relative deviation from \(\widehat{\nu_{\max}}\) as per \citet{yu18}. Astroconformer's relative median absolute error for the `Detection' group stands at 2.14\%, compared with the 2.52\% of the conventional pipeline from \citet{Hekker_2011}. Remarkably, even in the `Non-detection' subgroup where \citet{Hekker_2011} fails to identify oscillation modes, \textit{Astroconformer} maintains a relative median absolute error of 2.91\%.

\begin{figure*}
    \centering
    \includegraphics[width=1.0\textwidth]{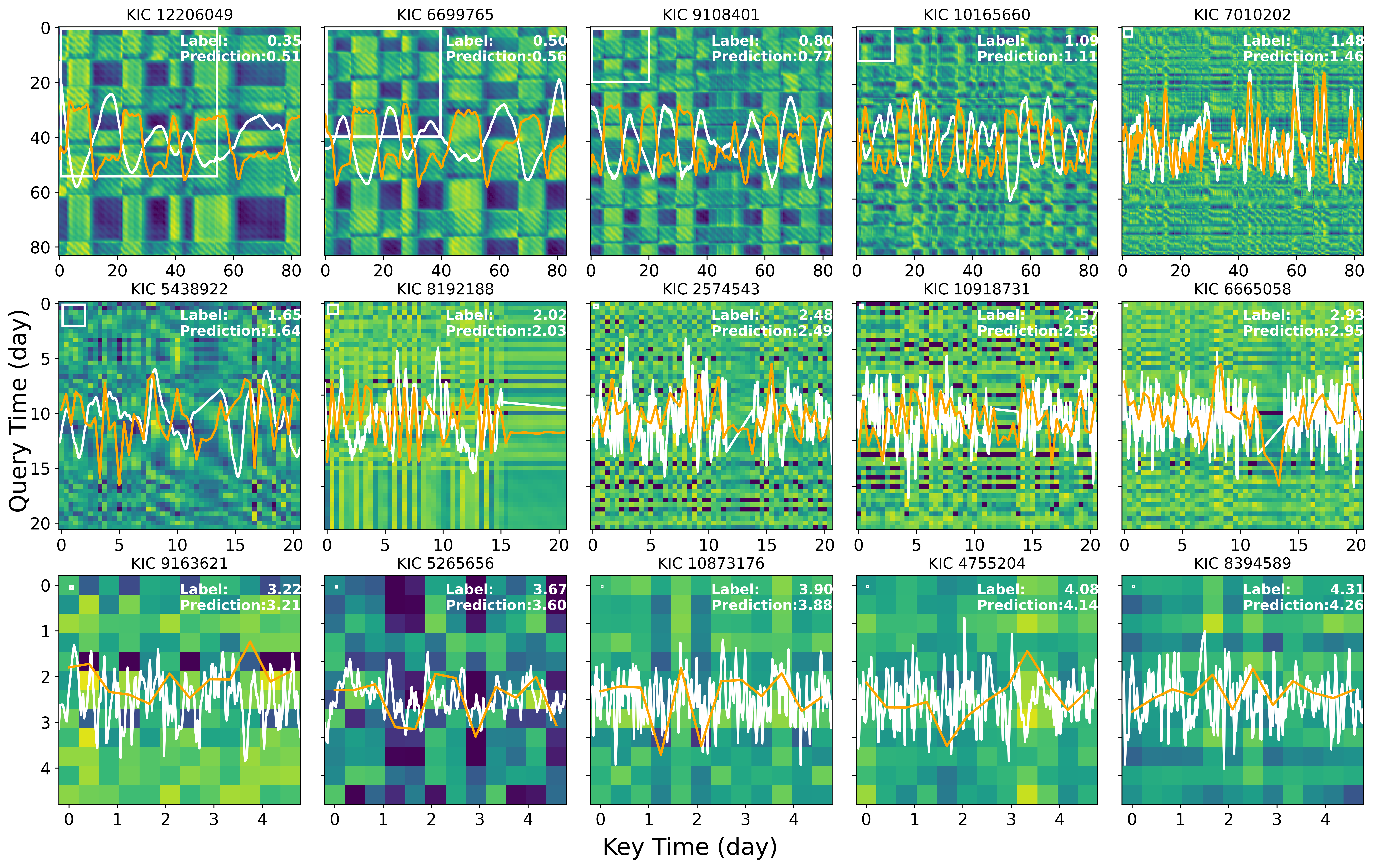}
    \caption{Attention maps for a selected set of \textit{Kepler} stars that span a broad range of \( \logg \) values and encompass various evolutionary phases, from turnoff stars to high-luminosity red giants. The brighter color indicate a larger attention weight. The star's normalized light curve is included for contextual reference. The row-averaged attention map are in orange to investigate the general pattern. Shown on top of each panel is the KIC number. A white square in the top-left corner of each panel indicates the star's characteristic oscillation period, as derived from \citet{yu18}. For better details, panels in the second row and third row are zoomed in by 4 and 16, respectively. For stars with lower \( \logg \) values (\( \logg < 2 \)), the periodic patterns in the attention maps closely match the stars' light curve waveforms, highlighting \textit{Astroconformer}'s proficiency in leveraging oscillations and granulation signals to accurately estimate \( \logg \). Interestingly, even for stars with higher \( \logg \) values—where one would expect the oscillation period to be unresolved in the attention map—long-range patterns remain discernible. These patterns have periods significantly longer than those of the stellar oscillations, further confirming that \textit{Astroconformer} is capable of capturing features associated with larger-period granulation, particularly in stars with higher \( \logg \) values.}
    \label{fig:attentionmap}
\end{figure*}

Additionally, \textit{Astroconformer} exhibits exceptional performance for high-luminosity red giants with \(\nu_{\max}<10\mu\mathrm{Hz}\), an achievement that is noteworthy given the challenges the automated pipeline faces with these stars. The primary challenge lies in the short timespan of light curves, which reduces the frequency resolution in power spectra, thereby complicating the detection and characterization of oscillations.  This effect is significant even when stellar oscillation are detectable, resulting in numerous outliers in \(\nu_{\max}\) estimates by the automated pipeline, especially for faint red giants with \(\nu_{\max}>100\mu\mathrm{Hz}\) whose oscillation amplitudes are lower. In contrast, \textit{Astroconformer} demonstrates remarkable resilience against such intricate cases, further evidencing its robustness.

%
%
%
%
%
%
\section{Discussion}
\label{sec:discussion}

In this study, we demonstrated that \textit{Astroconformer} is able to provide precise and consistent results when analyzing single-quarter light curves. We also confirm its potential to provide a robust oscillation parameter estimation based on 30-day Kepler light curves. This performance advantage is likely attributable to \textit{Astroconformer}'s ability to ingest the entire time-series data, thereby capturing not only stellar oscillation patterns but also granulation, which are typically discarded in traditional pipelines. Moreover, \textit{Astroconformer}'s direct learning from light curves allows it to account for non-Gaussian phase information that is absent in power spectra. This capability is hypothesized to stem from \textit{Astroconformer}'s unique facility for capturing both local and global correlations within the data.

Supporting this hypothesis, our ablation study indicates that the \textit{Astroconformer} block employs an attention mechanism to calculate correlations between timestamp patches. To delve deeper into \textit{Astroconformer}'s inner workings, we intend to visualize these learned attention patterns below.

\subsection{Understanding Astroconformer's Learning Process}
\label{sec:attentionmap}

One of the distinct benefits of the self-attention mechanism is its capacity for global analysis, which contrasts with the localized receptive fields in CNNs. Furthermore, the self-attention framework provides an intuitive framework for interpreting what the neural network has learned \citep{dino}. As outlined in Section~\ref{sec:selfattn}, the key operation in self-attention involves taking the dot product of key and query vectors, which are linear transformations of the original input. The softmax of this product yields an attention score, quantifying the similarity between two patches in the input sequence. This attention matrix offers a valuable perspective into the specific aspects of the light curve that \textit{Astroconformer} prioritizes.

In particular, the input of \textit{Astroconformer} is 200 patches, each contains 20 timestamps (10 hours), of the light curve. In self-attention mechanism, each patch is converted to a query patch and a key patch. The key patch at position  $j$ is paired with the query patch at position $i$, yielding their correlation, denoted as $A_{ij} = softmax(\boldsymbol{q}^i\cdot\boldsymbol{k}^j)$. Subsequently, these correlations serve as weights of how the patch $i$ combines information from all 200 patches. 

With this in mind, in Figure~\ref{fig:attentionmap}, we display attention maps for a diverse set of \textit{Kepler} stars, spanning a range of \( \logg \) values and stellar types, from turnoff stars to high-luminosity red giants. For the sake of illustration, we examine the attention map generated by the first head of the MHSA module within \textit{Astroconformer}'s third layer. The background image of each map shows the attention values for individual patches, indicating how correlated signals are at different moments in time. Overlaying each map in the top-left corner is a white square, which signifies the star's characteristic oscillation period. Additionally, the star's normalized light curve is included for contextual reference. For the readability of general attention trend, a row-averaged attention map is plotted against each light curve. Lastly, we note that diagonal stripes in the attention maps are artifacts of the positional encoding process, since it scales the correlation of position $i, j$ by a complex phase proportional to their relative distance $i-j$.

First, as seen in the figure, \textit{Astroconformer}'s attention mechanism operates across varied time intervals to enable a more comprehensive analysis of light curves directly in the time domain. This is in contrast with CNNs, which are limited by their local receptive fields, making them less adaptable for analyzing extended time series data like light curves. Consequently, CNN approaches in asteroseismology are often restricted to the scrutiny of power spectra \citep[e.g.,][]{Hon2018}. For stars with \( \logg < 2 \) that the general attention patterns generated by \textit{Astroconformer} are consistent with the light curve stellar oscillations. But on top of the characteristic oscillation periods, the patterns also include low frequency components, potentially from granulation.

Equally illuminating are the attention maps for stars with \( \logg > 3 \). These maps highlight frequency below those typically associated with oscillations. Given that these maps examine 10-hour patches—exceeding the oscillation periods of stars with \( \logg > 3 \)—one would expect any oscillation patterns to remain unresolved. If stellar oscillations were the only feature captured, the attention maps would offer limited insights. Yet, as Figure~\ref{fig:attentionmap} illustrates, averaged attention maps show consistency with  trends of light curves. These extended patterns, substantially longer than the oscillation periods, support the idea that \textit{Astroconformer} is tapping into additional data layers, possibly related to granulation and the non-gaussian nature of oscillations, to accurately infer the \( \logg \) of stars. We also further validate the correlation between attention maps and granulation timescales for red giants in Appendix~\ref{sec:attnmap_gran}.

Finally, another significant observation relates to the interpolated segments within the light curve. In these areas, the attention map typically assigns lower weights, thereby minimizing the influence of interpolated information. For instance, in the case of KIC 6665058 (\( \logg \approx 3 \)), an interpolated segment around the 15th day appears notably dimmer in the attention map compared to surrounding areas. This suggests that \textit{Astroconformer} demonstrates resilience against contaminated signals, effectively distinguishing spurious data from genuine light curve features. This capability may also account for \textit{Astroconformer}'s robust performance.

\subsection{Limitations and Future Work}
\label{sec:limitation}

While we have demonstrated that \textit{Astroconformer} holds considerable potential, there remain limitations that necessitate additional research. A fundamental constraint is our reliance on uniformly sampled light curves. While this is a reasonable assumption for data from missions like \textit{Kepler} or TESS, it could present challenges when dealing with non-uniform datasets, such as those generated by the Rubin Observatory. Despite the model's ability to leverage long-range temporal information, its current implementation may not be directly applicable to non-uniform data sources.

This assumption of uniform sampling is critical for two primary reasons. First, directly inputting individual time stamps without patching into the \textit{Astroconformer} block would significantly increase computational costs. For example, this approach would necessitate performing 4000 x 4000 attention calculations instead of a more manageable 200 x 200 grid. Second, the assumption aids the integration of convolutional modules within the \textit{Astroconformer} architecture. As discussed in Section~\ref{sec:vscnn}, the limited availability of training samples makes a stronger inductive bias via convolutional modules advantageous. While switching to a pure Transformer architecture could bypass the need for uniform sampling, it may also result in compromised model performance. Future work should investigate alternative methods for embedding or patching to accommodate non-uniformly sampled light curves.

Additionally, the ability of \textit{Astroconformer} to process all available information including instrumental effects simultaneously presents both advantages and disadvantages. Although the model can effectively filter out brief instances of spurious data, as evidenced in our empirical study in Section~\ref{sec:attentionmap}, it is susceptible to retaining misleading trends that persist throughout the entire light curve. Our empirical findings further substantiate this observation. For instance, we examine outliers in Section~\ref{sec:vsswan} and identify anomalies in their light curves, which are attributed to factors such as contamination, rotation modulation, or incomplete removal of extreme flux variations. Moreover, when applied to \texttt{Kepseismic} light curves\footnote{\texttt{Kepseismic} data are available at MAST via \url{http://dx.doi.org/10.17909/t9-mrpw-gc07}
}, where only signals with periods longer than 80 days are filtered and significant residual trends remain, \textit{Astroconformer}'s performance deteriorates, yielding an error of approximately 0.14 dex. This underscores the importance of meticulous data preprocessing, as outlined in Section~\ref{sec:data}. 
We consider this a limitation that could potentially be mitigated through deep learning techniques, such as constrastive learning \citep{simclr}, that allow for better data representation and spurious trend removal.

In this research, we have utilized training labels derived from the end-of-mission Kepler light curves in \citet{yu18}. These labels are of relatively low uncertainty, especially when compared to the inherent systematics present in inference models, including our own \textit{Astroconformer}. Therefore, they are not the dominant factor in this study. However, it is important to acknowledge that these labels are not free from noise. In future studies, exploring a more sophisticated probabilistic deep learning approach that accounts for label noise could be highly beneficial. Such an approach would aim to refine the inference process by effectively `deconvolving' the label noise, thereby enhancing the accuracy and reliability of the results.

%
%
%
%
%
%
\section{Conclusions}
\label{sec:conclusions}

In this study, we introduce \textit{Astroconformer}, a novel method based on the Transformer neural network architecture, renowned for its capacity to adaptively extract long-range contextual information from sequential data. This is particularly suited for time-series data such as stellar light curves, which often exhibit long-range dependencies. We specifically employ \textit{Astroconformer} to infer the surface gravities of stars from their light curves. The primary findings of our research can be summarized as follows:

\begin{itemize}
\item[1.] \textit{Astroconformer} excels in the precise inference of \( \logg \) values. When benchmarked against asteroseismic \( \logg \) measurements, spanning a \( \logg \) range of 0.2--4.4 dex, \textit{Astroconformer} outperforms the \( k \)-nearest neighbor-based method known as {\sc the swan}. \textit{Astroconformer} achieves a minimal RMSE of 0.017 dex at \(\logg\sim3\) and presents an increase in RMSE, up to 0.1 dex, towards the boundaries of the investigated \(\logg\) range.

\item[2.] The efficacy of a deep learning model in extracting \( \logg \) values from light curves is highly dependent on the model's receptive field. When self-attention modules are replaced with convolutional modules, there is a marked decline in performance, underscoring the superiority of Transformer models over convolutional neural networks for handling stellar light curves.

\item[3.] When predicting $\nu_{\max}$ from single-quarter light curves from \textit{Kepler}, \textit{Astroconformer} achieves a relative median absolute error of 1.93\% and a relative median standard deviation of 2.10\%, compared with the SYD pipeline's relative median absolute error of 2.10\% and relative median standard deviation of 3.37\%.

\item[4.] For shorter 33-day light curves from Kepler, our analysis reveals that \textit{Astroconformer} attains a relative median absolute error of less than 3\%. It consistently delivers reliable \(\nu_{\max}\) estimations in nearly all instances. Conversely, the conventional pipeline fails to detect solar-like oscillations in approximately 30\% of these samples, highlighting \textit{Astroconformer}'s robust capabilities to infer $\nu_{\max}$, especially for short time series.

\item[5.] \textit{Astroconformer} also provides valuable interpretability via its attention maps. The attention maps show that the model has learned to identify not only stellar oscillations but also longer-period granulation patterns, which are crucial for making accurate \( \logg \) estimations, especially for less evolved stars.
\end{itemize}

While Transformer-based models have achieved considerable advancements in areas such as computer vision and natural language processing, their application in astronomy is still in its early stages. This study zeros in on applying Transformer models to stellar light curves, demonstrating their potential to better extract long-range information—information that has often been overlooked by CNNs. This capability could be particularly crucial in upcoming surveys like those from the Rubin Observatory and the Roman Space Telescope, as well as in extracting valuable stellar properties from data with more complex noise characteristics like TESS.

Looking at the broader picture, the utility of Transformer models undoubtedly extends well beyond these specific applications. Beyond the realm of asteroseismology, Transformer architectures hold immense potential for a wide array of other astronomical investigations. Given their proficiency in processing long-range information—a ubiquitous trait in astronomical data—these models open up avenues for more synergistic progress between the domains of machine learning and astronomy.

\section*{Acknowledgements}

We extend our gratitude to Maryum Sayeed and Saskia Hekker for generously sharing the data used in their work and for providing meticulous guidance on their utilization. YST acknowledges financial support received from the Australian Research Council via the DECRA Fellowship, grant number DE220101520. We also recognize the invaluable contribution of public data from the \textit{Kepler} mission, supplied by NASA's Ames Research Center. The \textit{Kepler} mission is funded by NASA's Science Mission Directorate. Our special thanks go to the \textit{Kepler} team for making their data openly accessible, thereby facilitating this research.

\section*{Data Availability}

The data used in this study are publicly available and can be accessed through the Kepler Public Archive hosted by the Mikulski Archive for Space Telescopes (MAST). Our PyTorch implementation of \textit{Astroconformer} is publicly available at \url{https://github.com/panjiashu/Astroconformer}. Any additional data or materials related to this paper may be requested from the corresponding author.


\bibliographystyle{mnras}
\bibliography{astroconformer}

\begin{thebibliography}{}
\makeatletter
\relax
\def\mn@urlcharsother{\let\do\@makeother \do\$\do\&\do\#\do\^\do\_\do\%\do\~}
\def\mn@doi{\begingroup\mn@urlcharsother \@ifnextchar [ {\mn@doi@} {\mn@doi@[]}}
\def\mn@doi@[#1]#2{\def\@tempa{#1}\ifx\@tempa\@empty \href {http://dx.doi.org/#2} {doi:#2}\else \href {http://dx.doi.org/#2} {#1}\fi \endgroup}
\def\mn@eprint#1#2{\mn@eprint@#1:#2::\@nil}
\def\mn@eprint@arXiv#1{\href {http://arxiv.org/abs/#1} {{\tt arXiv:#1}}}
\def\mn@eprint@dblp#1{\href {http://dblp.uni-trier.de/rec/bibtex/#1.xml} {dblp:#1}}
\def\mn@eprint@#1:#2:#3:#4\@nil{\def\@tempa {#1}\def\@tempb {#2}\def\@tempc {#3}\ifx \@tempc \@empty \let \@tempc \@tempb \let \@tempb \@tempa \fi \ifx \@tempb \@empty \def\@tempb {arXiv}\fi \@ifundefined {mn@eprint@\@tempb}{\@tempb:\@tempc}{\expandafter \expandafter \csname mn@eprint@\@tempb\endcsname \expandafter{\@tempc}}}

\bibitem[\protect\citeauthoryear{Aerts}{Aerts}{2021}]{astero_interior}
Aerts C.,  2021, \mn@doi [Reviews of Modern Physics] {10.1103/revmodphys.93.015001}, 93

\bibitem[\protect\citeauthoryear{Aerts, Christensen-Dalsgaard  \& Kurtz}{Aerts et~al.}{2010}]{astero1}
Aerts C.,  Christensen-Dalsgaard J.,   Kurtz D.,  2010, Asteroseismology.
Springer, Netherlands, \mn@doi{10.1007/978-1-4020-5803-5}

\bibitem[\protect\citeauthoryear{Auvergne et~al.,}{Auvergne et~al.}{2009}]{corot}
Auvergne M.,  et~al., 2009, Astronomy \& Astrophysics, 506, 411

\bibitem[\protect\citeauthoryear{Bedding et~al.,}{Bedding et~al.}{2011}]{astero_evo}
Bedding T.~R.,  et~al., 2011, \mn@doi [Nature] {10.1038/nature09935}, 471, 608

\bibitem[\protect\citeauthoryear{Bellm et~al.,}{Bellm et~al.}{2018}]{ztf}
Bellm E.~C.,  et~al., 2018, \mn@doi [Publications of the Astronomical Society of the Pacific] {10.1088/1538-3873/aaecbe}, 131, 018002

\bibitem[\protect\citeauthoryear{Benomar, Appourchaux  \& Baudin}{Benomar et~al.}{2009}]{gaussian2}
Benomar O.,  Appourchaux T.,   Baudin F.,  2009, \mn@doi [http://dx.doi.org/10.1051/0004-6361/200911657] {10.1051/0004-6361/200911657}, 506

\bibitem[\protect\citeauthoryear{Berger, Huber, van Saders, Gaidos, Tayar  \& Kraus}{Berger et~al.}{2020}]{berger}
Berger T.~A.,  Huber D.,  van Saders J.~L.,  Gaidos E.,  Tayar J.,   Kraus A.~L.,  2020, The Gaia-Kepler Stellar Properties Catalog. I. Homogeneous Fundamental Properties for 186,301 Kepler Stars (\mn@eprint {arXiv} {2001.07737})

\bibitem[\protect\citeauthoryear{{Blancato}, {Ness}, {Huber}, {Lu}  \& {Angus}}{{Blancato} et~al.}{2022}]{cnn_lc}
{Blancato} K.,  {Ness} M.~K.,  {Huber} D.,  {Lu} Y.,   {Angus} R.,  2022, \mn@doi [\apj] {10.3847/1538-4357/ac7563}, \href {https://ui.adsabs.harvard.edu/abs/2022ApJ...933..241B} {933, 241}

\bibitem[\protect\citeauthoryear{{Brown}, {Gilliland}, {Noyes}  \& {Ramsey}}{{Brown} et~al.}{1991}]{astero_pmode}
{Brown} T.~M.,  {Gilliland} R.~L.,  {Noyes} R.~W.,   {Ramsey} L.~W.,  1991, \mn@doi [\apj] {10.1086/169725}, \href {https://ui.adsabs.harvard.edu/abs/1991ApJ...368..599B} {368, 599}

\bibitem[\protect\citeauthoryear{Brown et~al.,}{Brown et~al.}{2020}]{gpt3}
Brown T.~B.,  et~al., 2020, Language Models are Few-Shot Learners (\mn@eprint {arXiv} {2005.14165})

\bibitem[\protect\citeauthoryear{Bugnet, Garc{\'{\i} }a, Davies, Mathur, Corsaro, Hall  \& Rendle}{Bugnet et~al.}{2018}]{fliper}
Bugnet L.,  Garc{\'{\i} }a R.~A.,  Davies G.~R.,  Mathur S.,  Corsaro E.,  Hall O.~J.,   Rendle B.~M.,  2018, \mn@doi [A{\&}A] {10.1051/0004-6361/201833106}, 620, A38

\bibitem[\protect\citeauthoryear{Caron, Touvron, Misra, Jégou, Mairal, Bojanowski  \& Joulin}{Caron et~al.}{2021}]{dino}
Caron M.,  Touvron H.,  Misra I.,  Jégou H.,  Mairal J.,  Bojanowski P.,   Joulin A.,  2021, Emerging Properties in Self-Supervised Vision Transformers (\mn@eprint {arXiv} {2104.14294})

\bibitem[\protect\citeauthoryear{{Chaplin} \& {Miglio}}{{Chaplin} \& {Miglio}}{2013}]{astero_damp}
{Chaplin} W.~J.,  {Miglio} A.,  2013, \mn@doi [\araa] {10.1146/annurev-astro-082812-140938}, \href {https://ui.adsabs.harvard.edu/abs/2013ARA&A..51..353C} {51, 353}

\bibitem[\protect\citeauthoryear{Chatterjee \& Zielinski}{Chatterjee \& Zielinski}{2022}]{generalization}
Chatterjee S.,  Zielinski P.,  2022, On the Generalization Mystery in Deep Learning (\mn@eprint {arXiv} {2203.10036})

\bibitem[\protect\citeauthoryear{Chen, Kornblith, Norouzi  \& Hinton}{Chen et~al.}{2020}]{simclr}
Chen T.,  Kornblith S.,  Norouzi M.,   Hinton G.,  2020, A Simple Framework for Contrastive Learning of Visual Representations (\mn@eprint {arXiv} {2002.05709})

\bibitem[\protect\citeauthoryear{Devlin, Chang, Lee  \& Toutanova}{Devlin et~al.}{2019}]{bert}
Devlin J.,  Chang M.-W.,  Lee K.,   Toutanova K.,  2019, BERT: Pre-training of Deep Bidirectional Transformers for Language Understanding (\mn@eprint {arXiv} {1810.04805})

\bibitem[\protect\citeauthoryear{Dosovitskiy et~al.,}{Dosovitskiy et~al.}{2021}]{vit}
Dosovitskiy A.,  et~al., 2021, An Image is Worth 16x16 Words: Transformers for Image Recognition at Scale (\mn@eprint {arXiv} {2010.11929})

\bibitem[\protect\citeauthoryear{{Dupret} et~al.,}{{Dupret} et~al.}{2009}]{astero_lifetime}
{Dupret} M.~A.,  et~al., 2009, \mn@doi [\aap] {10.1051/0004-6361/200911713}, \href {https://ui.adsabs.harvard.edu/abs/2009A&A...506...57D} {506, 57}

\bibitem[\protect\citeauthoryear{Erickson, Mueller, Shirkov, Zhang, Larroy, Li  \& Smola}{Erickson et~al.}{2020}]{autogluon}
Erickson N.,  Mueller J.,  Shirkov A.,  Zhang H.,  Larroy P.,  Li M.,   Smola A.,  2020, arXiv preprint arXiv:2003.06505

\bibitem[\protect\citeauthoryear{Fuller, Cantiello, Stello, Garcia  \& Bildsten}{Fuller et~al.}{2015}]{astero_magnetic1}
Fuller J.,  Cantiello M.,  Stello D.,  Garcia R.~A.,   Bildsten L.,  2015, \mn@doi [Science] {10.1126/science.aac6933}, 350, 423

\bibitem[\protect\citeauthoryear{Garc{\'{\i}}a \& Ballot}{Garc{\'{\i}}a \& Ballot}{2019}]{astero_solar}
Garc{\'{\i}}a R.~A.,  Ballot J.,  2019, \mn@doi [Living Reviews in Solar Physics] {10.1007/s41116-019-0020-1}, 16

\bibitem[\protect\citeauthoryear{Garc{\'{\i} }a et~al.,}{Garc{\'{\i} }a et~al.}{2011}]{astero_preprocess}
Garc{\'{\i} }a R.~A.,  et~al., 2011, \mn@doi [Monthly Notices of the Royal Astronomical Society: Letters] {10.1111/j.1745-3933.2011.01042.x}, 414, L6

\bibitem[\protect\citeauthoryear{{Gehan}, {Mosser}, {Michel}, {Samadi}  \& {Kallinger}}{{Gehan} et~al.}{2018}]{astero_rotation3}
{Gehan} C.,  {Mosser} B.,  {Michel} E.,  {Samadi} R.,   {Kallinger} T.,  2018, \mn@doi [\aap] {10.1051/0004-6361/201832822}, \href {https://ui.adsabs.harvard.edu/abs/2018A&A...616A..24G} {616, A24}

\bibitem[\protect\citeauthoryear{Gruberbauer, Kallinger, Weiss  \& Guenther}{Gruberbauer et~al.}{2009}]{gaussian3}
Gruberbauer M.,  Kallinger T.,  Weiss W.~W.,   Guenther D.~B.,  2009, \mn@doi [A{\&}A] {10.1051/0004-6361/200811203}, 506, 1043

\bibitem[\protect\citeauthoryear{Gulati et~al.,}{Gulati et~al.}{2020}]{conformer}
Gulati A.,  et~al., 2020, Conformer: Convolution-augmented Transformer for Speech Recognition (\mn@eprint {arXiv} {2005.08100})

\bibitem[\protect\citeauthoryear{{Hall} et~al.,}{{Hall} et~al.}{2021}]{astero_rotation2}
{Hall} O.~J.,  et~al., 2021, \mn@doi [Nature Astronomy] {10.1038/s41550-021-01335-x}, \href {https://ui.adsabs.harvard.edu/abs/2021NatAs...5..707H} {5, 707}

\bibitem[\protect\citeauthoryear{He, Zhang, Ren  \& Sun}{He et~al.}{2015}]{resnet}
He K.,  Zhang X.,  Ren S.,   Sun J.,  2015, Deep Residual Learning for Image Recognition (\mn@eprint {arXiv} {1512.03385})

\bibitem[\protect\citeauthoryear{Hekker \& Christensen-Dalsgaard}{Hekker \& Christensen-Dalsgaard}{2017}]{astero_giant}
Hekker S.,  Christensen-Dalsgaard J.,  2017, \mn@doi [The Astronomy and Astrophysics Review] {10.1007/s00159-017-0101-x}, 25

\bibitem[\protect\citeauthoryear{Hekker et~al.,}{Hekker et~al.}{2011}]{Hekker_2011}
Hekker S.,  et~al., 2011, \mn@doi [Monthly Notices of the Royal Astronomical Society] {10.1111/j.1365-2966.2011.18574.x}, 414, 2594

\bibitem[\protect\citeauthoryear{Hekker et~al.,}{Hekker et~al.}{2012}]{Hekker_2012}
Hekker S.,  et~al., 2012, \mn@doi [A{\&}A] {10.1051/0004-6361/201219328}, 544, A90

\bibitem[\protect\citeauthoryear{{Hinners}, {Tat}  \& {Thorp}}{{Hinners} et~al.}{2018}]{rnn_lc}
{Hinners} T.~A.,  {Tat} K.,   {Thorp} R.,  2018, \mn@doi [\aj] {10.3847/1538-3881/aac16d}, \href {https://ui.adsabs.harvard.edu/abs/2018AJ....156....7H} {156, 7}

\bibitem[\protect\citeauthoryear{Hon, Stello  \& Yu}{Hon et~al.}{2018}]{Hon2018}
Hon M.,  Stello D.,   Yu J.,  2018, \mn@doi [Monthly Notices of the Royal Astronomical Society] {10.1093/mnras/sty483}, 476, 3233–3244

\bibitem[\protect\citeauthoryear{Hon, Stello, Garc{\'{\i} }a, Mathur, Sharma, Colman  \& Bugnet}{Hon et~al.}{2019}]{non-solar}
Hon M.,  Stello D.,  Garc{\'{\i} }a R.~A.,  Mathur S.,  Sharma S.,  Colman I.~L.,   Bugnet L.,  2019, \mn@doi [Monthly Notices of the Royal Astronomical Society] {10.1093/mnras/stz622}, 485, 5616

\bibitem[\protect\citeauthoryear{Howell et~al.,}{Howell et~al.}{2014}]{K2}
Howell S.~B.,  et~al., 2014, \mn@doi [Publications of the Astronomical Society of the Pacific] {10.1086/676406}, 126, 398

\bibitem[\protect\citeauthoryear{Huber, Stello, Bedding, Chaplin, Arentoft, Quirion  \& Kjeldsen}{Huber et~al.}{2009}]{astero_syd}
Huber D.,  Stello D.,  Bedding T.~R.,  Chaplin W.~J.,  Arentoft T.,  Quirion P.-O.,   Kjeldsen H.,  2009, Automated extraction of oscillation parameters for Kepler observations of solar-type stars (\mn@eprint {arXiv} {0910.2764})

\bibitem[\protect\citeauthoryear{Ioffe \& Szegedy}{Ioffe \& Szegedy}{2015}]{batchnorm}
Ioffe S.,  Szegedy C.,  2015, Batch Normalization: Accelerating Deep Network Training by Reducing Internal Covariate Shift (\mn@eprint {arXiv} {1502.03167})

\bibitem[\protect\citeauthoryear{Ismail~Fawaz, Forestier, Weber, Idoumghar  \& Muller}{Ismail~Fawaz et~al.}{2019}]{tsc}
Ismail~Fawaz H.,  Forestier G.,  Weber J.,  Idoumghar L.,   Muller P.-A.,  2019, \mn@doi [Data Mining and Knowledge Discovery] {10.1007/s10618-019-00619-1}, 33, 917–963

\bibitem[\protect\citeauthoryear{{Ivezi{\'c}} et~al.,}{{Ivezi{\'c}} et~al.}{2019}]{lsst}
{Ivezi{\'c}} {\v{Z}}.,  et~al., 2019, \mn@doi [\apj] {10.3847/1538-4357/ab042c}, \href {https://ui.adsabs.harvard.edu/abs/2019ApJ...873..111I} {873, 111}

\bibitem[\protect\citeauthoryear{Kallinger et~al.,}{Kallinger et~al.}{2014}]{granulation_oscillation}
Kallinger T.,  et~al., 2014, \mn@doi [A{\&}A] {10.1051/0004-6361/201424313}, 570, A41

\bibitem[\protect\citeauthoryear{{Kallinger}, {Hekker}, {Garcia}, {Huber}  \& {Matthews}}{{Kallinger} et~al.}{2016}]{logg_timescale}
{Kallinger} T.,  {Hekker} S.,  {Garcia} R.~A.,  {Huber} D.,   {Matthews} J.~M.,  2016, \mn@doi [Science Advances] {10.1126/sciadv.1500654}, \href {https://ui.adsabs.harvard.edu/abs/2016SciA....2E0654K} {2, 1500654}

\bibitem[\protect\citeauthoryear{Kawaler \& Hostler}{Kawaler \& Hostler}{2005}]{astero_rotation1}
Kawaler S.~D.,  Hostler S.~R.,  2005, \mn@doi [The Astrophysical Journal] {10.1086/427403}, 621, 432

\bibitem[\protect\citeauthoryear{{Kjeldsen} \& {Bedding}}{{Kjeldsen} \& {Bedding}}{1995}]{astero_numax}
{Kjeldsen} H.,  {Bedding} T.~R.,  1995, \mn@doi [\aap] {10.48550/arXiv.astro-ph/9403015}, \href {https://ui.adsabs.harvard.edu/abs/1995A&A...293...87K} {293, 87}

\bibitem[\protect\citeauthoryear{{Koch} et~al.}{{Koch} et~al.}{2010}]{Kepler}
{Koch} D.~G.,  et~al., 2010, \mn@doi [\apjl] {10.1088/2041-8205/713/2/L79}, \href {https://ui.adsabs.harvard.edu/abs/2010ApJ...713L..79K} {713, L79}

\bibitem[\protect\citeauthoryear{{Li}, {Deheuvels}, {Ballot}  \& {Ligni{\`e}res}}{{Li} et~al.}{2022}]{astero_magnetic2}
{Li} G.,  {Deheuvels} S.,  {Ballot} J.,   {Ligni{\`e}res} F.,  2022, \mn@doi [\nat] {10.1038/s41586-022-05176-0}, \href {https://ui.adsabs.harvard.edu/abs/2022Natur.610...43L} {610, 43}

\bibitem[\protect\citeauthoryear{Loshchilov \& Hutter}{Loshchilov \& Hutter}{2019}]{adamw}
Loshchilov I.,  Hutter F.,  2019, Decoupled Weight Decay Regularization (\mn@eprint {arXiv} {1711.05101})

\bibitem[\protect\citeauthoryear{Marshall et~al.,}{Marshall et~al.}{2017}]{lsst_strategy}
Marshall P.,  et~al., 2017, LSST Science Collaborations Observing Strategy White Paper: "Science-driven Optimization of the LSST Observing Strategy", \mn@doi{10.5281/ZENODO.842713}, \url {https://zenodo.org/record/842713}

\bibitem[\protect\citeauthoryear{{Mathur} et~al.,}{{Mathur} et~al.}{2011}]{granulation_acoustic}
{Mathur} S.,  et~al., 2011, \mn@doi [\apj] {10.1088/0004-637X/741/2/119}, \href {https://ui.adsabs.harvard.edu/abs/2011ApJ...741..119M} {741, 119}

\bibitem[\protect\citeauthoryear{Mathur et~al.,}{Mathur et~al.}{2017}]{mathur}
Mathur S.,  et~al., 2017, \mn@doi [The Astrophysical Journal Supplement Series] {10.3847/1538-4365/229/2/30}, 229, 30

\bibitem[\protect\citeauthoryear{Ness, Aguirre, Lund, Cantiello, Foreman-Mackey, Hogg  \& Angus}{Ness et~al.}{2018}]{ness2018}
Ness M.~K.,  Aguirre V.~S.,  Lund M.~N.,  Cantiello M.,  Foreman-Mackey D.,  Hogg D.~W.,   Angus R.,  2018, \mn@doi [The Astrophysical Journal] {10.3847/1538-4357/aadb40}, 866, 15

\bibitem[\protect\citeauthoryear{Nie, Nguyen, Sinthong  \& Kalagnanam}{Nie et~al.}{2023}]{patchtst}
Nie Y.,  Nguyen N.~H.,  Sinthong P.,   Kalagnanam J.,  2023, A Time Series is Worth 64 Words: Long-term Forecasting with Transformers (\mn@eprint {arXiv} {2211.14730})

\bibitem[\protect\citeauthoryear{Ramachandran, Zoph  \& Le}{Ramachandran et~al.}{2017}]{swish}
Ramachandran P.,  Zoph B.,   Le Q.~V.,  2017, Searching for Activation Functions (\mn@eprint {arXiv} {1710.05941})

\bibitem[\protect\citeauthoryear{{Ricker} et~al.,}{{Ricker} et~al.}{2014}]{tess}
{Ricker} G.~R.,  et~al., 2014, in {Oschmann} Jacobus~M. J.,  {Clampin} M.,  {Fazio} G.~G.,   {MacEwen} H.~A.,  eds,  Society of Photo-Optical Instrumentation Engineers (SPIE) Conference Series Vol. 9143, Space Telescopes and Instrumentation 2014: Optical, Infrared, and Millimeter Wave. p. 914320 (\mn@eprint {arXiv} {1406.0151}), \mn@doi{10.1117/12.2063489}

\bibitem[\protect\citeauthoryear{Sayeed, Huber, Wheeler  \& Ness}{Sayeed et~al.}{2021}]{swan}
Sayeed M.,  Huber D.,  Wheeler A.,   Ness M.~K.,  2021, \mn@doi [The Astronomical Journal] {10.3847/1538-3881/abdf4c}, 161, 170

\bibitem[\protect\citeauthoryear{Smith}{Smith}{2017}]{cyclic}
Smith L.~N.,  2017, Cyclical Learning Rates for Training Neural Networks (\mn@eprint {arXiv} {1506.01186})

\bibitem[\protect\citeauthoryear{Stello, Cantiello, Fuller, Huber, Garc{\'{\i}}a, Bedding, Bildsten  \& Aguirre}{Stello et~al.}{2016}]{astero_magnetic3}
Stello D.,  Cantiello M.,  Fuller J.,  Huber D.,  Garc{\'{\i}}a R.~A.,  Bedding T.~R.,  Bildsten L.,   Aguirre V.~S.,  2016, \mn@doi [Nature] {10.1038/nature16171}, 529, 364

\bibitem[\protect\citeauthoryear{Su, Lu, Pan, Wen  \& Liu}{Su et~al.}{2021}]{RoPE}
Su J.,  Lu Y.,  Pan S.,  Wen B.,   Liu Y.,  2021, RoFormer: Enhanced Transformer with Rotary Position Embedding (\mn@eprint {arXiv} {2104.09864})

\bibitem[\protect\citeauthoryear{Szegedy et~al.,}{Szegedy et~al.}{2014}]{googlenet}
Szegedy C.,  et~al., 2014, Going Deeper with Convolutions (\mn@eprint {arXiv} {1409.4842})

\bibitem[\protect\citeauthoryear{{Toutain} \& {Appourchaux}}{{Toutain} \& {Appourchaux}}{1994}]{gaussian1}
{Toutain} T.,  {Appourchaux} T.,  1994, \aap, \href {https://ui.adsabs.harvard.edu/abs/1994A&A...289..649T} {289, 649}

\bibitem[\protect\citeauthoryear{{Ulrich}}{{Ulrich}}{1986}]{astero_dn}
{Ulrich} R.~K.,  1986, \mn@doi [\apjl] {10.1086/184700}, \href {https://ui.adsabs.harvard.edu/abs/1986ApJ...306L..37U} {306, L37}

\bibitem[\protect\citeauthoryear{Vaswani, Shazeer, Parmar, Uszkoreit, Jones, Gomez, Kaiser  \& Polosukhin}{Vaswani et~al.}{2017}]{attentionisallyouneed}
Vaswani A.,  Shazeer N.,  Parmar N.,  Uszkoreit J.,  Jones L.,  Gomez A.~N.,  Kaiser L.,   Polosukhin I.,  2017, Attention Is All You Need (\mn@eprint {arXiv} {1706.03762})

\bibitem[\protect\citeauthoryear{Wang, Ma, Dong, Huang, Zhang  \& Wei}{Wang et~al.}{2022}]{deepnorm}
Wang H.,  Ma S.,  Dong L.,  Huang S.,  Zhang D.,   Wei F.,  2022, DeepNet: Scaling Transformers to 1,000 Layers (\mn@eprint {arXiv} {2203.00555})

\bibitem[\protect\citeauthoryear{Yu, Huber, Bedding, Stello, Hon, Murphy  \& Khanna}{Yu et~al.}{2018}]{yu18}
Yu J.,  Huber D.,  Bedding T.~R.,  Stello D.,  Hon M.,  Murphy S.~J.,   Khanna S.,  2018, \mn@doi [The Astrophysical Journal Supplement Series] {10.3847/1538-4365/aaaf74}, 236, 42

\bibitem[\protect\citeauthoryear{Yu, Bedding, Stello, Huber, Compton, Gizon  \& Hekker}{Yu et~al.}{2020}]{Yu20}
Yu J.,  Bedding T.~R.,  Stello D.,  Huber D.,  Compton D.~L.,  Gizon L.,   Hekker S.,  2020, \mn@doi [Monthly Notices of the Royal Astronomical Society] {10.1093/mnras/staa300}, 493, 1388

\makeatother
\end{thebibliography}


\appendix
\section{Training}
\label{sec:train}

In Section~\ref{sec:vsswan} and Section~\ref{sec:vscnn}, the 5-layer \textit{Astroconformer} and other models are trained with the AdamW optimizer \citep{adamw} and used a cyclic learning rate scheduler \citep{cyclic}. This scheduler features three linear warm-up and cool-down stages (warm-up ratio=0.1), with each subsequent stage having a peak learning rate that is half of its predecessor. The lower bound of learning rate is set to be 0.1\% of the first peak value. We set the first peak learning rate to 0.003 such that the model is converged while fully trained given the other setting, except for `Without MHSA', which adopts a 0.001 learning rate due to training instability. Regarding the batch size selection, we conducted tests with batch sizes of 128, 256, and 512. Our findings indicated that batch size does not significantly impact the final result, provided it is greater than 256. Consequently, we opted for a batch size of 256 for our experiments. Training was carried out over 20,000 steps, with validation loss results indicating that this step count is adequate; we observed a plateau in the validation loss beyond this point. The typical training duration for \textit{Astroconformer} on an Nvidia A100 40GB GPU is approximately 3.5 hours. 

In Section~\ref{sec:vssyd}, the training of a 20-layer \textit{Astroconformer} involves weight initialization scheme proposed by \citet{deepnorm}, which sets the shortcut weight to 2 and weight initialization gain to 0.5. We then implement gradient clipping by 1.0. The model was trained 100,000 training steps, with a batch size of 48, a warm-up stage lasting 10,000 steps, a peak learning rate of 0.0003, and a lower bound learning rate, 10\% of the peak value. The training takes about 50 hours. In Section~\ref{sec:vshekker}, the finetuning of the 20-layer \textit{Astroconformer} contains a warm-up and cool-down stage lasting 10,000 steps, with a peak learning rate of 0.0001.

\section{Relation between Attention Maps and Granulation}
\label{sec:attnmap_gran}

We hypothesize that \textit{Astroconformer}'s robust performance compared to conventional pipelines is due to its ability to extract information from granulation within the light curves, in addition to focusing on oscillation modes. This is supported by observations in Figure~\ref{fig:attentionmap}, where the representation of high \(\logg\) values seems to incorporate more than just oscillation modes, suggesting that additional data, including granulation, contributes to the inference of \(\logg\) for dwarf stars. However, it's important to acknowledge that while attention maps offer an intuitive insight into the workings of the attention mechanism, they do not directly link to granulation information. To clarify this connection, we aim to demonstrate a relationship between the patterns observed in Figure~\ref{fig:attentionmap} and granulation timescales measured by the SYD pipeline. Specifically, the SYD pipeline fits the power spectrum background with the form

\begin{equation}
    P(\nu)=P_{n}+\sum_{i=1}^{k} \frac{4 \sigma_{i}^{2} \tau_{i}}{1+\left(2 \pi \nu \tau_{i}\right)^{2}+\left(2 \pi \nu \tau_{i}\right)^{4}},
\end{equation}
where \(P_n\) is the white noise component, \(k\) is the number of power laws used, and \(\sigma\) and \(\tau\) are the root-mean-square intensity and timescale of granulation, respectively \citep{astero_syd}. The timescale of the first power law component, $\tau_1$, is used to characterize the granulation timescale.

To explore the general response trends in attention maps, we randomly selected 3,487 single-quarter light curves from our dataset, as discussed in Section~\ref{sec:vsastero}. For each of these light curves, we calculated the attention map using the first head in the third self-attention layer and then computed the power spectrum for each row of the attention map. An averaged power spectrum over all rows is utilized as our input variable. Meanwhile, we run the SYD pipeline on these light curves, and the derived $\tau_1$ serves as our target variable. If the attention maps are indeed capturing granulation information, it should be feasible to deduce granulation timescales from the power spectra of these attention maps. However, this non-linear regression task presents a wide array of methodological options.

To streamline this analysis, we employ Autogluon \citep{autogluon}, an auto-ML package that systematically applies various machine learning models and combines their results. This approach enables us to effectively learn the correlation between the averaged power spectra and the granulation timescales. The 3,487 samples are split into subgroups of 2487, 500, 500 for training, validation, and test set, respectively.

We adopt the default \texttt{Tabularpredictor} in Autugluon to fit our training data, which automatically search through the methods including K-Nearest Neighbor, Light Gradient Boosting Machine (LightGBM), Random Forest, Categorical Boosting , Extra Trees, Multi Layer Perceptron, and eXtreme Gradient Boost (XGBoost), as well as optimising their parameters. Each model is evaluated by its RMSE on the validation set. After training individual models, an ensemble model is derived by weighting predictions from individual models by their RMSE on the validation set. In our experiment, the LightGBM model with an RMSE of 0.0098 day is the most precise individual model, while the ensemble model achieves a better RMSE of 0.0094 day.

\begin{figure}
    \centering
    \includegraphics[width=0.5\textwidth]{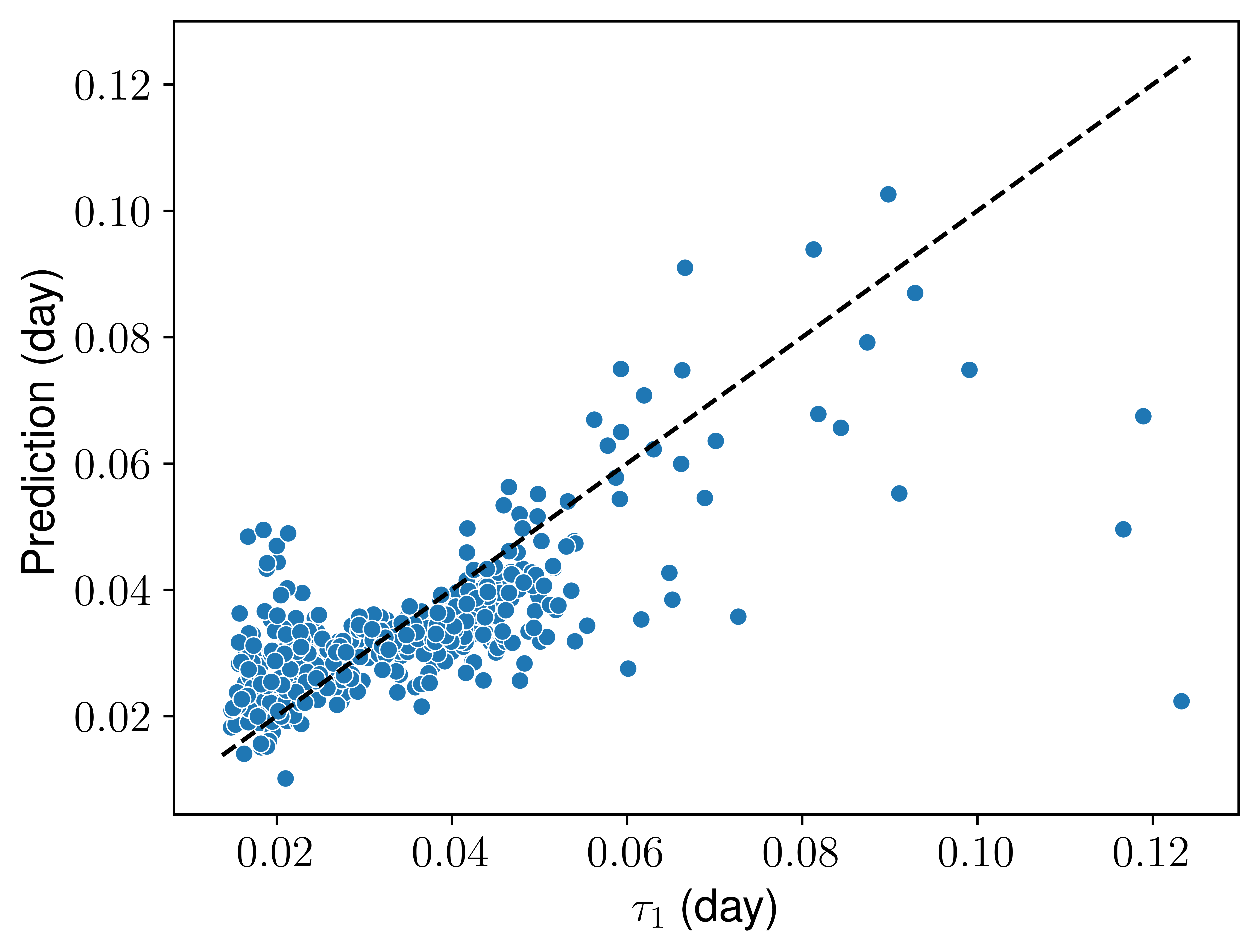}
    \caption{Averaged power spectra of attention maps encode granulation information. The horizontal axis is characteristic timescale of the first power law component of granulation, \(\tau_1\), derived by the SYD pipeline, and the vertical axis is \(\tau_1\) predicted by AutoML using averaged power spectra of attention maps.}
    \label{fig:attnmap_gran}
\end{figure}

We use the ensemble model to predict \(\tau_1\) based on averaged power spectra of attention maps in test set. The outcomes of our analysis are presented in Fig~\ref{fig:attnmap_gran}, with an RMSE of 0.0104 day. While this does not conclusively establish causality, this initial investigation into the power spectrum of the attention map lends further credence to the idea that the attention mechanism, as applied to light curves, indeed captures information reflective of the granulation patterns in stars. This evidence supports our assertion that {\it Astroconformer} excels at extracting insights beyond just oscillation modes.

\bsp	
\label{lastpage}
\end{CJK*}
\end{document}